\documentclass[twocolumn]{aastex63}

\def\arcsec{$^{\prime\prime}$} 
\newcommand{\msun}{M$_{\sun}$}
\newcommand{\rsun}{R$_{\sun}$}
\newcommand{\rstar}{R$_{\star}$}
\newcommand{\mstar}{M$_{\star}$}
\newcommand{\lsun}{L$_{\sun}$}
\newcommand{\msunyr}{\msun\,yr$^{-1}$}

\newcommand{\teff}{\rm $T_{eff}$}
\newcommand{\escm}{{\rm erg/s/cm}$^2$}
\newcommand{\halpha}{H$\alpha$}
\newcommand{\hbeta}{H$\beta$}
\newcommand{\brgamma}{Br$\gamma$}
\newcommand{\mdot}{$\dot{M}$}
\newcommand{\ri}{R$_{\rm i}$}

\def\curf{{\cal F}}
\newcommand{\vsini}{$v\sin i$}
\newcommand{\hst}{\textit{HST}}
\newcommand{\tess}{\textit{TESS}}
\newcommand{\lfl}{{\rm $\lambda F_{\lambda}$}}

\newcommand{\rin}{$R_{\rm i}$}

\newcommand{\av}{$A_V$}

\newcommand\be {\begin{equation}}
\newcommand\en{\end{equation}}
\def\micron{$\mu$m}

\def\fp1{$f_{p1}$}

\usepackage[utf8]{inputenc}
\usepackage{graphicx}
\usepackage{amsmath,amssymb}
\usepackage{breqn}
\usepackage{booktabs}
\usepackage{longtable}
\usepackage{lipsum}
\usepackage{amsmath}

\shorttitle{ODYSSEUS study of Orion OB1b}
\shortauthors{Pittman et al.}

\graphicspath{{./}{figures/}}

\begin{document}

\title{Towards a comprehensive view of accretion, inner disks, and extinction in classical T Tauri stars: an ODYSSEUS study of the Orion OB1b association}

\correspondingauthor{Caeley Pittman}
\email{cpittman@bu.edu}

\author[0000-0001-9301-6252]{Caeley V. Pittman}
\affiliation{Institute for Astrophysical Research, Department of Astronomy, Boston University, 725 Commonwealth Avenue, Boston, MA 02215, USA}

\author[0000-0001-9227-5949]{Catherine C. Espaillat}
\affiliation{Institute for Astrophysical Research, Department of Astronomy, Boston University, 725 Commonwealth Avenue, Boston, MA 02215, USA}

\author[0000-0003-1639-510X]{Connor E. Robinson}
\affiliation{Department of Physics \& Astronomy, Amherst College, Amherst, MA 01002, USA}

\author[0000-0003-4507-1710]{Thanawuth Thanathibodee}
\affiliation{Institute for Astrophysical Research, Department of Astronomy, Boston University, 725 Commonwealth Avenue, Boston, MA 02215, USA}

\author[0000-0002-3950-5386]{Nuria Calvet}
\affiliation{Department of Astronomy, University of Michigan, 311 West Hall, 1085 S. University Avenue, Ann Arbor, MI 48109, USA}

\author[0000-0002-6808-4066]{John Wendeborn}
\affiliation{Institute for Astrophysical Research, Department of Astronomy, Boston University, 725 Commonwealth Avenue, Boston, MA 02215, USA}

\author[0000-0001-9797-5661]{Jesus Hern{\'a}ndez}
\affil{Instituto de Astronom\'{i}a, Universidad Aut\'{o}noma de M\'{e}xico Ensenada, B.C, M\'{e}xico}

\author[0000-0003-3562-262X]{Carlo F. Manara}
\affiliation{European Southern Observatory, Karl-Schwarzschild-Strasse 2, 85748 Garching, Germany}

\author[0000-0001-7796-1756]{Fred Walter}
\affiliation{Department of Physics and Astronomy, Stony Brook University, Stony Brook NY 11794-3800, USA}

\author[0000-0001-6015-646X]{P\'eter \'Abrah\'am}
\affiliation{Konkoly Observatory, Research Centre for Astronomy and Earth Sciences, E\"otv\"os Lor\'and Research Network, Konkoly-Thege Mikl\'os \'ut 15-17, 1121 Budapest, Hungary}
\affiliation{ELTE E\"otv\"os Lor\'and University, Institute of Physics, P\'azm\'any P\'eter S\'et\'any 1/A, 1117 Budapest, Hungary}
\affiliation{CSFK, MTA Centre of Excellence, Konkoly-Thege Mikl\'os \'ut 15-17, 1121 Budapest, Hungary}

\author[0000-0001-8657-095X]{Juan M. Alcal\'a}
\affiliation{Osservatorio Astronomico di Capodimonte, via Moiariello 16, 80131 Napoli, Italy}

\author[0000-0002-5171-8376]{S\'ilvia H. P. Alencar}
\affiliation{Departamento de Fisica, Universidade Federal de Minas Gerais, Av. Antonio Carlos 6627, 30270-901 Belo Horizonte, MG, Brazil}

\author[0000-0003-2631-5265]{Nicole Arulanantham}
\affiliation{Space Telescope Science Institute, 3700 San Martin Drive, Baltimore, MD 21218, USA}

\author[0000-0002-1593-3693]{Sylvie Cabrit}
\affiliation{Observatoire de Paris, PSL University, Sorbonne University, CNRS, LERMA, 61 Av. de l'Observatoire, 75014 Paris, France}

\author[0000-0001-6496-0252]{Jochen Eisl\"offel}
\affiliation{Th\"uringer Landessternwarte, Sternwarte 5, D-07778 Tautenburg, Germany}

\author[0000-0002-5261-6216]{Eleonora Fiorellino}
\affiliation{Konkoly Observatory, Research Centre for Astronomy and Earth Sciences, E\"otv\"os Lor\'and Research Network, Konkoly-Thege Mikl\'os \'ut 15-17, 1121 Budapest, Hungary}
\affiliation{CSFK, MTA Centre of Excellence, Konkoly-Thege Mikl\'os \'ut 15-17, 1121 Budapest, Hungary}
\affiliation{Osservatorio Astronomico di Capodimonte, via Moiariello 16, 80131 Napoli, Italy}

\author[0000-0002-1002-3674]{Kevin France}
\affiliation{Laboratory for Atmospheric and Space Physics, University of Colorado Boulder, Boulder, CO 80303, USA}

\author[0000-0002-8364-7795]{Manuele Gangi}
\affiliation{Osservatorio Astronomico di Roma, via di Frascati 33, 00078 Monte Porzio Catone, Italy}

\author[0000-0001-5707-8448]{Konstantin Grankin}
\affiliation{Crimean Astrophysical Observatory, Department of Stellar Physics, Nauchny, 298409, Crimea}

\author[0000-0002-7154-6065]{Gregory J. Herczeg}
\affiliation{Kavli Institute for Astronomy and Astrophysics, Peking University, Yiheyuan 5, Haidian Qu, 100871 Beijing, China}
\affiliation{Department of Astronomy, Peking University, Yiheyuan 5, Haidian Qu, 100871 Beijing, China}

\author[0000-0001-7157-6275]{\'Agnes K\'osp\'al}
\affiliation{Konkoly Observatory, Research Centre for Astronomy and Earth Sciences, E\"otv\"os Lor\'and Research Network, Konkoly-Thege Mikl\'os \'ut 15-17, 1121 Budapest, Hungary}
\affiliation{ELTE E\"otv\"os Lor\'and University, Institute of Physics, P\'azm\'any P\'eter S\'et\'any 1/A, 1117 Budapest, Hungary}
\affiliation{CSFK, MTA Centre of Excellence, Konkoly-Thege Mikl\'os \'ut 15-17, 1121 Budapest, Hungary}
\affiliation{Max Planck Institute for Astronomy, K\"onigstuhl 17, 69117 Heidelberg, Germany}

\author[0000-0002-0233-5328]{Ignacio Mendigut\'\i{}a}
\affil{Centro de Astrobiolog\'\i{}a (CSIC-INTA), ESA-ESAC Campus, 28692, Villanueva de la Cañada, Madrid, Spain}

\author[0000-0001-7351-6540]{Javier Serna}
\affil{Instituto de Astronom\'{i}a, Universidad Aut\'{o}noma de M\'{e}xico Ensenada, B.C, M\'{e}xico}

\author[0000-0002-4115-0318]{Laura Venuti}
\affiliation{SETI Institute, 339 Bernardo Ave, Suite 200, Mountain View, CA 94043, USA}

\submitjournal{AJ}
\received{15 June 2022}
\revised{2 August 2022}
\accepted{7 August 2022}

\keywords{Stellar accretion disks; Protoplanetary disks; Star formation; Pre-main sequence stars; Reddening law; Interstellar extinction}

\begin{abstract}

The coevolution of T Tauri stars and their surrounding protoplanetary disks dictates the timescales of planet formation. In this paper, we present magnetospheric accretion and inner disk wall model fits to NUV-NIR spectra of nine classical T Tauri stars in Orion OB1b as part of the {\it Outflows and Disks around Young Stars: Synergies for the Exploration of ULLYSES Spectra} (ODYSSEUS) Survey. Using NUV-optical spectra from the {\it Hubble UV Legacy Library of Young Stars as Essential Standards} (ULLYSES) Director's Discretionary Program and optical-NIR spectra from the PENELLOPE VLT Large Programme, we find that the accretion rates of these targets are relatively high for the region's intermediate age of 5.0~Myr; rates range from 0.5--$17.2 \times 10^{-8}$ \msunyr, with a median value of $1.2\times 10^{-8}$ \msunyr. 
The NIR excesses can be fit with 1200--1800~K inner disk walls located at 0.05--0.10~AU from the host stars. We discuss the significance of the choice in extinction law, as the measured accretion rate depends strongly on the adopted extinction value. This analysis will be extended to the complete sample of T Tauri stars being observed through ULLYSES to characterize accretion and inner disks in star-forming regions of different ages and stellar populations.
\end{abstract}

\section{Introduction}
\label{sec:intro}
The coevolution of T Tauri stars (TTS) and their surrounding protoplanetary disks is one of the most important subjects in the field of planet formation \citep{manara22}. Accretion of disk material onto the star can be traced by continuum excesses in the near-ultraviolet to optical region of the spectrum \citep{valenti93,gullbring00,hh08}. Emission from the frontally-illuminated inner wall of the disk can be traced by continuum excesses in the near-to mid-infrared region \citep{natta01,tuthill01,dalessio05}. A model that can self-consistently reproduce the spectra of actively-accreting T Tauri stars from the ultraviolet to the infrared is a vital tool for understanding the coevolution of the star and inner disk. To reproduce the NUV-NIR spectra of CTTS, one can combine accretion shock and inner disk wall models to create a consistent description of the physical mechanisms producing the observed spectra, as has been demonstrated in GM Aurigae by \cite{ingleby15}.

The magnetospheric accretion model \citep{koenigl91, shu94, hartmann94, cg98,muzerolle98,muzerolle01} has been widely used in explaining observations of UV-excesses and emission lines in CTTS \citep[for a review, see][]{hartmann16}. \cite{bouvier20} and the \cite{gravity20} confirmed that hydrogen \brgamma\ emission associated with the magnetospheric accretion paradigm originates well within TTS corotation radii, reinforcing the hypothesis that the emission comes from accretion columns rather than a wind farther away from the star. When modeling this emission, it is important to fit the optical spectra in addition to the NUV, as accretion estimates that fit excesses only at blue wavelengths underestimate the accretion rates by a factor of $\sim$2 \citep{fischer11}. To account for both NUV and optical continuum excesses in classical T Tauri stars (CTTS), \cite{ingleby13} found that a multi-column accretion shock model should be used \citep[based on the shock model of][]{cg98}.

\cite{dalessio05} found that the spectrum of a heated inner wall of a dusty disk located at the dust sublimation radius dominates the emission at NIR-MIR wavelengths. The structure and intensity of emission from the inner wall of protoplanetary disks are determined primarily by the wall's geometry and the mineralogy of the dust \citep[e.g.][]{muzerolle03,mcclure13}. Gas accretion onto the star is likely driven by turbulence in this inner disk, which also necessarily causes diffusion of dust particles that are coupled to the gas; the strength of this diffusion dictates the shape of the inner wall, independent of dust composition, with the wall geometry ranging from short and curved, to tall and vertical \citep{schobert21}. The primary implication of this is that observed emission from a curved wall does not depend heavily on disk inclination $i$, whereas emission from a vertical wall depends strongly on $i$.
\cite{mcclure13} best fit the 2--10~\micron\ continuum excesses of four TTS by approximating a curved wall using two vertical walls at different radii, each with a different height, dust size distribution, and composition.

The \textit{Hubble UV Legacy Library of Young Stars as Essential Standards} (ULLYSES) DDT program \citep{ullyses20}, combined with the PENELLOPE Very Large Telescope (VLT) Large Program \citep{manara21}, provides an ideal sample for exploring the variation in CTTS system properties across ranges of spectral types, ages, and masses. The {\it Outflows and Disks around Young Stars: Synergies for the Exploration of ULLYSES Spectra} (ODYSSEUS) collaboration \citep{espaillat22} is maximizing the scientific impact of these data by studying accretion, outflows, and disk chemistry in the largest sample of TTS observed with the {\it Hubble Space Telescope} (\hst) to date. This paper focuses on modeling the accretion and inner disks of the first sample of TTS observed by ULLYSES.

ULYSSES began observations with Orion OB1, located about 400~pc away \citep{briceno19}. Here, we study nine CTTS in the Orion OB1b star-forming region, which has a mean age of 5.0 Myr and a disk fraction of 12\% \citep{briceno19}. This is intermediate between Cloud B (2.5 Myr, 31\% CTTS) and OB1a (10.8 Myr, 6\% CTTS), indicating an intermediate level of disk evolution in the region.

In this work, we model the same Orion OB1b sample of CTTS as \cite{manara21} to demonstrate the significance of using NUV spectra to determine the accretion and extinction properties of these targets. For the first time, we model both accretion and inner disks in a sample of CTTS using contemporaneous spectra from 2000--24000~\AA. Once this analysis is extended to the full ULLYSES sample of $\sim$60 CTTS in nine star-forming regions, we can constrain the relationships between stellar parameters and accretion and inner disk properties to an extent that has not been done before.

This paper is organized as follows. In Section~\ref{sec:sample}, we discuss the sample of TTS, multiplicity in the sample, stellar parameters, and observations. In Section~\ref{sec:model}, we describe the accretion shock and disk models used to fit the NUV-NIR continua of the CTTS. Section~\ref{sec:results} presents the modeling results. In Section~\ref{sec:discussion}, we compare our results to those of previous studies and discuss the significance of the choice in extinction law and the wavelength range of data available. Finally, in Section~\ref{sec:summary} we present a summary of our findings and future work that will be possible through ODYSSEUS.

\section{Sample and Observations}
\label{sec:sample}
The ULLYSES program \citep{ullyses20} began observations with a sample of eight CTTS in the Orion OB1b subassociation \citep{briceno01,briceno05,briceno19} and two WTTS (weak-lined TTS, non-accreting sources) in the 25 Ori cluster, which are used as template photospheres in this analysis (see Figure~\ref{fig:17and36} in Appendix~\ref{sec:app_A} for a justification of their classification as WTTS). Each target was observed with both \hst\ and the VLT contemporaneously. One of these targets, CVSO~165, was discovered to be a binary system composed of two CTTS, so 11 TTS were observed in total spanning spectral types M3.5--K6 and masses 0.3--0.9~\msun.

\subsection{Objects and stellar parameters}
\label{sec:stellar_params}
CVSO~17, CVSO~36, and CVSO~109 are known to be physical binaries \citep{tokovinin20}, and CVSO~104 and CVSO~165 have visual companions with which they are not kinematically associated \citep{manara21}. The primary component of CVSO~104 was found to be a spectroscopic binary \citep{kounkel19,manara21,frasca21}, but these components are not resolved in the \hst\ observations. \cite{proffitt21} found that the primary component of CVSO~165 is also itself a binary. The ULLYSES \hst\ observations of this object are able to resolve the two components, which both show signatures of active accretion and are thus both modeled in this analysis.

Stellar parameters for all CTTS targets and WTTS photospheric templates used in this work are listed in Table~\ref{tab:stellar_params}. For targets not affected by unresolved binarity in their X-Shooter observations, values for spectral type (SpT), mass (\mstar), distance ($d$), and veiling ($r$) are adopted from \cite{manara21}, and their values for stellar radius (\rstar) and V-band extinction (\av) are included for comparison with the values we derive. Effective temperatures for all targets except the CVSO~165 binary system come from the temperature-spectral type relation given for 5--30 Myr stars in \cite{pm13}. \cite{manara21} derive distances from Gaia Early Data Release 3 parallaxes that have reliable astrometric solutions (typically RUWE $<$ 1.4). For those without reliable solutions, the mean distance to the association is used and an uncertainty of 10\% assumed.

\subsubsection{CVSO~165 binary} \label{sec:165}

We obtain the effective temperature (\teff) and the visual extinction  (\av) for each component of the CVSO~165 system by comparing the HST spectra against the PHOENIX synthetic spectral library \citep{husser13}. We use values of surface gravity adequate for PMS stars (e.g., $3.0<{\rm log(g)}<4.5$) and interpolate the spectral library to obtain theoretical grids with intervals of 50~K in {\teff}. For each theoretical spectrum, we apply values of {\av} from 0 to 5 in steps of 0.01 using the standard interstellar reddening law (R$_V$=3.1) from \citet{cardelli89}. Minimizing $\chi^2$, we obtain the {\teff} and the {\av} from the best match between the observed \hst\ spectra and the reddened synthetic spectra. To avoid contamination from accretion flux, we use the spectral range from 6000--9000 \AA. We also avoid the region around \halpha, fitting a Gaussian function to the line profile and ignoring the spectral range between -5$\sigma$ and +5$\sigma$ from the center of the line. Finally, the uncertainties in the estimated values are obtained using the Monte Carlo (MC) method of error propagation \citep{Anderson1976}, varying the \hst\ spectra fluxes randomly 500 times within their reported uncertainties. The values reported in Table~\ref{tab:stellar_params} are the median and standard deviation of the 500 MC results.  

We derive V magnitudes for each component of CVSO~165 by multiplying the \hst\ spectrum by the Johnson V-filter transmission curve. We obtained ${\rm V}=13.875\pm0.002$ for CVSO~165A and ${\rm V}=15.775\pm0.004$ for CVSO~165B. The RUWE astrometric parameter is highly sensitive to unresolved binary systems; the high RUWE value of 14.3 for CVSO~165 indicates a poor astrometric solution \citep{Lindegren2020}. Therefore, we assume a distance of $400\pm40$~pc \citep{briceno19} to estimate the stellar luminosity for CVSO~165A and CVSO~165B using the visual magnitude corrected by extinction. Finally, comparing the location on the H-R diagram with the MIST evolutionary model \citep{Dotter2016}, we obtained the stellar masses reported in Table~\ref{tab:stellar_params} for each component. The uncertainties in the masses were obtained using the MC method of error propagation and the uncertainties estimated for the luminosity and \teff.

\subsubsection{Veiling}
Veiling measurements are necessary to set the flux of the WTTS template photosphere relative to the CTTS spectrum, with the relationship given by $F_{{\rm phot},\lambda}=F_{{\rm CTTS},\lambda}/(1+r_\lambda)$. In our analysis, we scale the WTTS template spectrum to the data using the observed veiling at one wavelength, $\lambda_0$. First, the WTTS spectrum is scaled to the observed CTTS spectrum at $\lambda_0$ with a $\frac{F_{\rm obs}(\lambda _0)}{F_{\rm phot}(\lambda _0)}$ term. Then, it is scaled to the observed veiling with a $\frac{1}{1+r_{\lambda _0}}$ term such that the contribution of the photosphere to the data's continuum flux at $\lambda_0$ is equivalent to the contribution implied by the veiling measurement. Only the absolute flux of the WTTS template spectrum is changed, not the shape of the WTTS spectrum. This scaling process allows us to use the observed veiling to determine the amount of continuum excess emission for which we need to account with the accretion shock and inner disk wall models.

Three epochs of veiling measurements at $\lambda_0$=5500~\AA\ (r$_{5500}$) are available for 7 out of 9 of the CTTS from VLT/UVES and VLT/ESPRESSO \citep{manara21}, and we use the value from the epoch closest in time to the \hst/STIS observations for these targets. CVSO~90 has VLT/X-Shooter veiling available at $\lambda_0$=7100~\AA, so this is used in place of r$_{5500}$ for this target. The veiling measured from the unresolved CVSO~165 spectrum is split between the two components according to the ratio of their U-band fluxes. CVSO~104 has not yet been formally modeled at the wavelengths relevant to accretion; however, its r$_{5500}$ has been estimated from modeling the UVES spectrum when the objects are nearly in conjunction (A. Frasca, priv. comm.). The observations used to measure the veiling for each target are shown in Table~\ref{tab:obs_log}, and the veiling values are included in Table~\ref{tab:stellar_params}. 

\begin{deluxetable}{lcc}
\tablecaption{Observation Log  \label{tab:obs_log}}
\tablehead{
\colhead{Object} & \colhead{Telescope/Instrument} & \colhead{Date (MJD)}
}
\startdata
CVSO~58 & \hst/STIS$^{a}$ & 59184.97 \\
 & VLT/X-Shooter$^{b}$ & 59185.25 \\
 & VLT/UVES$^{b,c}$ & 59185.09 \\
\midrule
CVSO~90 & \hst/STIS$^{a}$ & 59199.60 \\
 & VLT/X-Shooter$^{b,c}$ & 59198.06 \\
\midrule
CVSO~104 & \hst/STIS$^{a}$ & 59180.21 \\
 & VLT/X-Shooter$^{b}$ & 59180.13 \\
 & VLT/UVES$^{b,c}$ & 59179.12 \\
\midrule
CVSO~107 & \hst/STIS$^{a}$ & 59188.02 \\
 & VLT/X-Shooter$^{b}$ & 59187.08 \\
 & VLT/UVES$^{b,c}$ & 59188.14 \\
\midrule
CVSO~109 & \hst/STIS$^{a}$ & 59181.20 \\
 & VLT/X-Shooter$^{b}$ & 59181.15 \\
 & VLT/UVES$^{b,c}$ & 59181.17 \\
\midrule
CVSO~146 & \hst/STIS$^{a}$ & 59192.85 \\
 & VLT/X-Shooter$^{b}$ & 59192.08 \\
 & VLT/ESPRESSO$^{b,c}$ & 59193.09 \\
\midrule
CVSO~165 & \hst/STIS$^{a}$ & 59197.82 \\
 & VLT/X-Shooter$^{b}$ & 59197.08 \\
 & VLT/ESPRESSO$^{b,c}$ & 59198.10 \\
\midrule
CVSO~176 & \hst/STIS$^{a}$ & 59182.86 \\
 & VLT/X-Shooter$^{b}$ & 59185.17 \\
 & VLT/UVES$^{b,c}$ & 59183.15 \\
\enddata
\tablecomments{$^{a}$Observed through the ULLYSES \hst\ DDT Program \citep{ullyses20}. $^{b}$Observed through the PENELLOPE VLT Large Programme \citep{manara21}. $^{c}$Used for veiling measurement.}
\end{deluxetable}

\begin{deluxetable*}{lccccccccc}
\rotate
\tablecaption{Adopted Stellar Properties of Orion OB1 Targets  \label{tab:stellar_params}}
\tablehead{
\colhead{Object} & \colhead{SpT} & \colhead{Temperature} & \colhead{Luminosity} & \colhead{Radius} & \colhead{Mass} & \colhead{Distance} & \colhead{\av} & \colhead{$r$} & \colhead{Template} \\
 & & (K) & (\lsun) & (\rsun) & (\msun) & (pc) & (mag) & & }
\startdata
\multicolumn{10}{c}{Single CTTS} \\
\cmidrule(lr){2-9}
CVSO~58 & K7 & 3970 & 0.32 & 1.19 & 0.81 & 349.00$\pm2.8$ & 0.8 & $0.81\pm0.04$ & HBC 427+{\it TWA 6} \\
CVSO~90 & M0.5 & 3700 & 0.13 & 0.88 & 0.62 & 338.70$^{+3.8}_{-3.7}$ & 0.1 & $1.8\pm0.4^{*}$ & TWA 7+{\it TWA 14} \\ 
CVSO~107 & M0.5 & 3700 & 0.32 & 1.38 & 0.53 & 330.40$\pm2.5$ & 0.3 & $0.98\pm0.11$ & TWA 7+{\it TWA 14} \\ 
CVSO~146 & K6 & 4020 & 0.80 & 1.84 & 0.86 & 332.00$\pm1.7$ & 0.6 & $0.44\pm0.10$ & HBC 427+{\it RXJ1543.1-3920} \\
CVSO~176 & M3.5 & 3260 & 0.34 & 1.83 & 0.25 & 302.40$^{+2.9}_{-2.8}$ & 1.0 & $0.34\pm0.16$ & CVSO~17/36+{\it TWA 15A} \\
\midrule
\multicolumn{10}{c}{Binary CTTS} \\
\cmidrule(lr){2-9}
CVSO~104$^{\dagger}$ & M2 & 3490 & 0.37 & 1.66 & 0.37 & 360.70$^{+3.9}_{-3.8}$ & 0.2 & $0.8^{\star}$ & CVSO~17/36+{\it TWA 2A} \\
CVSO~109A$^{a}$ & M0$\pm0.5$ & 3767.6$\pm81.2$ & $0.59^{+0.17}_{-0.13}$ & $1.81\pm0.25$ & 0.50$_{-0.05}^{+0.07}$ & 400$\pm40$ & 0.06$_{-0.24}^{+0.24}$ & $0.90\pm0.08$ & TWA 7+{\it TWA 14} \\ 
CVSO~165A & K5.5$\pm1.0$ & 4221$\pm28$ & 0.90$_{-0.15}^{+0.19}$ & 1.78$\pm1.07$ & 0.84$\pm0.05$ & 400$\pm40$ & 0.32$\pm0.05$ & $0.21\pm0.03^{**}$ & RECX 1+{\it RXJ1543.1-3920} \\
CVSO~165B & M1$\pm1.5$ & 3849$\pm7$ & 0.47$_{-0.08}^{+0.10}$ & 1.55$\pm0.93$ & 0.58$\pm0.23$ & 400$\pm40$ & 1.35$\pm0.12$ & $0.15\pm0.02^{**}$ & TWA 7+{\it TWA 14} \\
\midrule
\multicolumn{10}{c}{WTTS} \\
\cmidrule(lr){2-9}
RECX 1$^{b}$ & K5 & 4140 & 1.0 & 1.8 & 0.9 & 97 & 0.0 & ... & ... \\
{\it RXJ1543.1-3920}$^{c}$ & K6 & 4020 & 0.40 & 1.19 & ... & 150 & $0.1\pm0.1$ & ... & ... \\
HBC 427$^{b}$ & K7 & 3970 & 0.8 & 1.9 & 0.8 & 140 & 0.0 & ... & ... \\
{\it TWA 6}$^{d}$ & K7 & 3970 & 0.11 & 0.67 & 0.66 & 51 & ... & ... & ... \\
{\it TWA 14}$^{d}$ & M0.5 & 3700 & 0.15 & 0.90 & 0.73 & 96 & ... & ... & ... \\
TWA 7$^{b}$ & M1 & 3630 & 0.5 & 1.8 & 0.5 & 50 & 0.0 & ... & ... \\
{\it TWA 2A}$^{d}$ & M2 & 3490 & 0.33 & 1.51 & 0.55 & 47 & ... & ... & ... \\
CVSO~17$^{\dagger}$ & M2 & 3490 & 0.30 & 1.50 & 0.37 & 414.2$^{+9.3}_{-8.9}$ & 0.0 & ... & ... \\
CVSO~36$^{\dagger}$ & M2 & 3490 & 0.22 & 1.28 & 0.39 & 335.5$\pm3.0$ & 0.1 & ... & ... \\
{\it TWA 15A}$^{d}$ & M3.5 & 3260 & 0.11 & 1.00 & 0.30 & 111 & ... & ... & ... \\
\enddata
\tablecomments{The following stellar parameters for the CTTS come from \cite{manara21} unless otherwise noted: spectral type, luminosity, mass, distance \citep[derived from GAIA Early Data Release 3 parallaxes when reliable solutions were available, see][]{gaia21}, \av, $r$. All veilings are at 5500~\AA\ unless otherwise noted. The adopted temperature comes from the temperature-spectral type relation for 5-30~Myr stars in \citet[][Table 6]{pm13}, with an average used for intermediate spectral types. The radius is calculated from the luminosity and temperature using the Stefan-Boltzmann relation. Italicized WTTS come from X-Shooter; otherwise, they come from \hst/STIS. The average of CVSO~17 and CVSO~36, weighted by their uncertainties, is used because the individual spectra have low SNR in the NUV. $^{a}$ \cite{espaillat22} $^{b}$ \cite{ingleby13} $^{c}$\citet{manara17} $^{d}$ \cite{manara13} \\
$^{\dagger}$Unresolved spectroscopic binary system \\
$^{*}$Veiling at 7100~\AA\ from X-Shooter \\
$^{**}$Veiling for the primary and secondary components of CVSO~165 attained by scaling the ESPRESSO veiling of $0.36\pm0.05$ for CVSO~165 to the ratio of the components' U-band fluxes \\
$^\star$ Veiling at 5500~\AA\ estimated from modeling the UVES spectrum when the binary components are nearly in conjunction (A. Frasca, priv. comm.)}
\end{deluxetable*}

\subsection{HST/STIS observations}
\label{sec:hst_obs}
\hst\ Space Telescope Imaging Spectrograph (STIS) and Cosmic Origins Spectrograph (COS) observations of these targets were taken as part of the ULLYSES DDT program through proposals GO16113, GO16114, and GO16115 (\citealt{2020RNAAS...4..205R}, PI: Julia Roman-Duval). This paper utilizes observations of each target with the following gratings: STIS/G230L (spectral resolution 500-1010, plate scale 0.025''/pixel, and NUV-MAMA pixel size of 25~\micron), STIS/G430L (spectral resolution 530-1040, plate scale 0.051''/pixel, and CCD pixel size of 21~\micron), and STIS/G750L (spectral resolution 530-1040, plate scale 0.051''/pixel, and CCD pixel size of 21~\micron), all using the 52X2 slit. These spectra span a total wavelength range of $1710-10,000$~\AA\ after they are combined and trimmed. See Table~\ref{tab:obs_log} for the times of observations. Values of \lfl\ that are less than $1\times 10^{-15}$~\escm\ are removed. The spectra analyzed here come from ULLYSES Data Release~4, which separates spectra for the resolvable binary systems (CVSO~36, CVSO~104, CVSO~109, and CVSO~165) into their constituent components. The data between 1710 and 3300~\AA\ are dereddened with our derived \av\ using the \cite{whittet04} extinction law based on HD 29647 in Taurus (normalized to the \cite{cardelli89} standard ISM law), as the \cite{whittet04} law removes a potential overcorrection of the 2175~\AA\ bump that is present in standard ISM extinction laws. From 3300~\AA\ onwards, the data are dereddened using the \cite{cardelli89} law to align with the analysis of \cite{manara21}. Both laws assume an interstellar reddening of R$_V=3.1$.

\subsection{VLT/X-Shooter observations}
\label{sec:xs_obs}
Contemporaneous VLT/X-Shooter observations were taken through the ESO PENELLOPE Large Programme \citep[][PI: Carlo Manara; see their Figures F.1 and F.2 for the X-Shooter data overplotted with the \hst\ data]{manara2021data}. See Table~\ref{tab:obs_log} for the times of observations. Though observations were taken in the UVB, VIS, and NIR arms, we use only the NIR spectra (spectral resolution $\sim$11600, plate scale 0.248''/pixel, using the 0.4'' slit) for modeling the inner disk wall because \hst\ data are available out to 1~\micron. As described in \cite{manara21}, these flux-calibrated data are dereddened using the \cite{cardelli89} extinction law and corrected for telluric absorption using {\it molecfit} v3.0.3 \citep{smette15, kausch15}. Again, points with \lfl $<1\times 10^{-15}$ \escm\ are removed. 

The X-Shooter data for the binary targets CVSO~109 and CVSO~165 are unresolved, so it is necessary to split the total flux between the two components to achieve a better representation of the resolved NIR spectra. As described in \citet{espaillat22}, the X-Shooter spectrum of CVSO~109A is scaled by the J-band fluxes of the two components and their difference found by \cite{tokovinin20}. The J-band flux difference is not available for CVSO~165A and CVSO~165B, so these spectra are simply scaled down to align with the \hst\ continua. This is just an approximation given that the two components have different spectral types. However, our inferred NIR flux ratio of the primary to the secondary of 2.4 is reasonably consistent with our measured V- and R-band flux ratios of 2.6 and 2.7, respectively.

The \hst\ and X-Shooter observations of CVSO~176 were separated by about 65 hours, and this seems to have produced a significant discontinuity between the data. This can likely be attributed to variability, as variations on the order of hours to days are expected from magnetosphere-disk interactions \citep{venuti17,sergison20,fischer22}. Both the flux and slope of the \hst\ and X-Shooter continua do not agree. However, if the X-Shooter spectrum is scaled up by a factor of 1.75 (corresponding to a change of 0.6 mag), it aligns with the \hst\ spectrum. This agrees with contemporaneous photometry from AAVSOnet that shows a decrease in Sloan $i$-band flux of 0.6 mag between MJD 9182.67--9189.84.\footnote{\url{http://www.astro.sunysb.edu/fwalter/SMARTS/Odysseus/cvso176.phot.html}} This likely indicates a change in the emission from the inner disk wall. Both the scaled and unscaled X-Shooter data are shown in the model fits to CVSO~109A, CVSO~165A, CVSO~165B, and CVSO~176 presented in Figure~\ref{fig:model_fits}.

\subsection{TESS observations} \label{sec:tess}
\tess, the Transiting Exoplanet Survey Satellite \citep{ricker14}, has observed these targets on two occasions, in Sector~6 (2018 Dec~11 through 2019 Jan~07) and in Sector~32 (2020 Nov~19 through Dec~17). The latter coincided with the observations reported here, and these data were used to estimate the stellar rotation periods of our targets \citep[expected of order 4--9 days,][]{percy10}. We note that the 27 day viewing window of {\it TESS} samples only a few of the expected periods, so this analysis only reveals gross trends in very complex light curves.

We download the full frame image data from the MAST archives using the TESScut
software \citep{brasseur19}. {\it TESS} images, while photometrically 
stable and of continuous cadence, suffer from coarse spatial resolution 
(21\arcsec\ pixels).
{\it TESS} is a single-channel photometer with a 600--1000~nm bandpass. 
The temporal resolution was 10~minutes in Sector~32 and 30~minutes in Sector~6.

We extract the data using aperture photometry with a 1.5 pixel radius. 
The background is extracted from an annulus between 5 and 10 pixels
from the source. Because there are often other sources in the background 
annulus, we iterate on the background pixels, removing those more than
3$\sigma$ from the median level until we converge on the median background
level. We assume that the background is spatially flat in this region.

\subsection{WTTS photospheric templates} \label{subsec:templates}
The template photospheres constructed for each of the CTTS targets are composed of two WTTS spectra stitched together: the wavelength range between 2000 and $\sim$6,000~\AA\ comes from the \textit{HST}/STIS spectrum closest in spectral type to the CTTS, and the remaining data out to 24,000~\AA\ come from the VLT/X-Shooter spectrum that provided the best photospheric fit in \cite{manara21} (except for CVSO~165B, which has a spectral type different than that assigned to the unresolved CVSO~165 system). All targets have \hst\ WTTS within $\pm1$ spectral subtype except for CVSO~176, which is fit with a WTTS 1.5 subtypes earlier. Each pair of WTTS stitched together as photospheric templates for each CTTS target are listed in the ``Template'' column of Table~\ref{tab:stellar_params}. Note that there are four X-Shooter WTTS for which extinction estimates are unavailable. \av\ for these targets is assumed to be zero because they were chosen from regions of low extinction \citep{manara13}. This should not have a significant effect on the fitted \av\ and \mdot\ values because extinction is most important at wavelengths shorter than these X-Shooter WTTS spectra cover.

\section{Accretion and Disk Models}
\label{sec:model}
The accretion and disk wall models used in this work are computed and fit in order to be consistent with one another. First the accretion shock model is calculated for the specific stellar parameters of each target and fit to the data. Then, the output stellar radius, accretion rate, and shock temperature are used as inputs to the disk wall model, which is then calculated for the given stellar parameters and fit to the data. In Sections~\ref{sec:shockmodel}--\ref{sec:diad}, we describe our implementation of the \cite{cg98} accretion shock model and the D'Alessio Irradiated Accretion Disk radiative transfer model (DIAD, \citealt{dalessio98, dalessio99, dalessio01, dalessio04, dalessio05, dalessio06}).

\subsection{Multi-column accretion shock model}
\label{sec:shockmodel}
We model the NUV and optical \hst\ continua using the \cite{cg98} accretion shock model, updated to include three accretion columns of varying energy fluxes. These approximate a flow with a density gradient, as was recently found in GM Aurigae \citep{espaillat21}. Following \cite{cg98}, we assume a magnetospheric truncation radius (\rin) of 5~\rstar\ for all objects but CVSO~109A.\footnote{Modeling of \halpha\ and \hbeta\ lines of CVSO~109A showed a smaller \rin\ of 2.3~\rstar\ \citep[see][]{espaillat22}, but this analysis is still in preparation for the other objects (Thanathibodee et al., in prep.). We note that changing \rin\ to 2.3~\rstar\ from 5~\rstar\ increases the calculated accretion rate by a factor of 1.4.} \rin, \rstar, and \mstar\ determine the infall velocity of the accreted material, which is assumed to be the freefall velocity. The updated model is solved for individual parameter combinations rather than interpolating over a presolved grid of solutions. We add V-band extinction \av\ as a free parameter in the Markov chain Monte Carlo (MCMC) fitting process, then calculate the stellar radius \rstar\ from the fluxes of the dereddened CTTS spectrum and its associated X-shooter WTTS template scaled by the measured veiling. For details regarding the updates applied to the model, see \cite{re19}.

The complete shock model is composed of: a WTTS template photosphere $F_{\rm phot}(\lambda)$ scaled to the veiling measured by \cite{manara21} $r_{\lambda _0}$ (shown in Table~\ref{tab:stellar_params}), plus three accretion columns of low, medium, and high flux densities ($\curf$ = $1\times10^{10}$, $1\times10^{11}$, and $1\times10^{12}$ erg s$^{-1}$ cm$^{-2}$, respectively). The emission from these columns is scaled by the stellar radius \rstar, the distance $d$, and filling factors $f$, which represent the fraction of the stellar surface that is covered by each accretion column. Thus, the total dereddened model flux is given by

{\small
\begin{equation} \label{eq:full_ftot}
    F_{\rm tot}(\lambda) = 10^{0.4A_{\lambda _0}}\left[\frac{F_{\rm obs}(\lambda _0)}{F_{\rm phot}(\lambda _0)}\frac{F_{\rm phot}(\lambda)}{1+r_{\lambda _0}}\right] + \left(\frac{R_\star}{d}\right)^2 \sum_{i}^{n}f_i\curf_i(\lambda)
\end{equation}}
\noindent where $A_{\lambda _0}$ is the extinction as described in Sections~\ref{sec:hst_obs} and \ref{sec:xs_obs}; $F_{\rm phot}(\lambda)$ is the WTTS template photosphere scaled by $F_{\rm obs}(\lambda _0)/F_{\rm phot}(\lambda _0)$ to the observed STIS spectrum at $\lambda _0$; $n$ sums over the three accretion columns of different energy-flux densities; $f_i$ are the filling factors associated with each accretion column; and $\curf_i$ are the accretion shock spectra calculated as the sum of the emission from the heated photosphere and the preshock region of the system for each accretion column.

Equation~\ref{eq:full_ftot} is fit to the \hst\ continuum (with emission lines masked out) between 2000-10,000~\AA\ for each object using an MCMC with 2,000 steps, 100 walkers, and a burn-in of 550. \av\ has a tophat prior ranging between 0 and 2 on account of the low extinction reported for Orion OB1 \citep[median $A_V=0.65$ mag, ][]{briceno19}, and the sum of the filling factors is restricted between 0 and 40\% of the stellar surface, as modeling has demonstrated that the footprint of the accretion column can produce detectable emission that covers up to 39\% of the TTS surface \citep{ingleby13,ingleby14,re19}. Once the best-fit model is obtained, the average temperature of the shock ($T_{\rm shock}$) is calculated by fitting blackbody curves to the three columns' spectra, then weighting each column's associated accretion luminosity by its fractional filling factor according to

{\small
\begin{equation} \label{eq:Tshock}
    T_{\rm shock} = \left[\frac{f_{1E10}}{f_{\rm tot}}T_{1E10}^4 + \frac{f_{1E11}}{f_{\rm tot}}T_{1E11}^4 + \frac{f_{1E12}}{f_{\rm tot}}T_{1E12}^4 \right]^{\frac{1}{4}}.
\end{equation}}

\noindent The blackbody given by $T_{\rm shock}$ is an important input to the inner disk wall model, as both the stellar and the accretion luminosities irradiate the wall.

\subsection{DIAD}
\label{sec:diad}
We also model the protoplanetary disk's frontally-illuminated inner dust wall, which is located at the dust sublimation radius, using the D'Alessio Irradiated Accretion Disk radiative transfer models (DIAD, \citealt{dalessio98, dalessio99, dalessio01, dalessio04, dalessio05, dalessio06}). The inner dust wall begins contributing significantly at 1~\micron, after which the total model consists primarily of emission from the photosphere and the disk, with non-zero but less significant emission from the accretion columns at these longer wavelengths. Because our data extend only to 2.4~\micron, our model includes only the inner wall, not the disk behind it.

For details regarding the DIAD model used here, see the ODYSSEUS I paper \citep[][Section~6.1.2]{espaillat22}. In short, we model the inner dust wall assuming a fractional abundance of graphite and pyroxine-type silicates of 0.0025 and 0.004, respectively, in accordance with the \citet{draine84} model for the diffuse ISM. The grains are spherical with a size distribution that scales as $a^{-p}$ between grain radii of $a_{\rm min}$ and $a_{\rm max}$ and $p$ of 3.5 \citep{mathis77}. The minimum grain size is held at 0.005~{\micron}. Since the NIR spectra do not extend to 10~{\micron}, we cannot confidently constrain the 10~{\micron} silicate feature. Instead, we assume an $a_{max}$ of 10~{\micron} because \citet{mauco18} fit the 10~{\micron} silicate feature well with this value for the three objects from this analysis that were included in their study (CVSO~104, CVSO~107, and CVSO~109). The wall is illuminated by the stellar luminosity and the accretion luminosity, which is given by the $T_{\rm shock}$ derived for each target from its best-fit accretion shock model.

To obtain the best fit to the SED of each target, we adjust the height of the inner wall ($z\rm _{wall}$) between 0.5 and 20 gas scale heights (H), noting that the larger values of $z\rm _{wall}$ are indicative of excess emission likely originating from an optically-thin dust cavity in a pre-transitional disk for which we do not account in this model \citep{mauco18}. We adjust the temperature of the optically thin wall atmosphere ($T\rm _{wall}$) between 1200 and 1800 K. Disk inclination $i$ is estimated as described in Section~\ref{sec:inclinations}.

Note that in the case of a vertical wall, the inner disk wall height is degenerate with the inclination of the system. Since these disks are unresolved, we cannot distinguish between a high wall and a highly inclined viewing angle (we receive maximum wall emission from a disk inclined at 60-80$^{\circ}$, see \citealt{dullemond01,calvet05}).

The radius in the disk at which the wall is located ($R_{\rm wall}$) is derived using the best-fitting $T_{\rm wall}$ following
\begin{equation}R_{\rm wall} \sim 
\left [{\frac{(L_* + L_{\rm acc})}{16 \pi \sigma_R} }
( 2 + { \frac{\kappa_s} {\kappa_d} }) \right ] ^ {1/2} 
{1 \over T_{\rm wall}^2 }, \label{eq:rwall}
\end{equation} 
\noindent which assumes that the thickness of the atmosphere is negligible compared to the radius \citep{muzerolle03,dalessio04}, where $\sigma_R$ is the Stefan-Boltzmann constant; $\kappa_s$ and $\kappa_d$ are the mean opacities to the incident and local radiation, respectively; $L_\star$ is the stellar luminosity; and $L_{\rm acc}$ is the luminosity of the stellar accretion shock as given by the output accretion rate (\mdot) and $R_\star$ of the multi-column accretion shock model described in Section~\ref{sec:shockmodel}, with
\begin{equation}
   L_{\rm acc} = (1 - 1/R_{\rm i})(GM_\star\dot{M}/R_\star).
\end{equation}

\noindent As described in Section~\ref{sec:shockmodel}, \rin\ is taken to be 2.3\rstar\ for CVSO~109A and 5\rstar\ for all other targets. Disk models with $R_{\rm wall}$ located at the dust sublimation radius predict values of $R_{\rm wall}$ between 0.07-0.54 AU, with stronger accretors having larger values of $R_{\rm wall}$ as indicated by Equation~\ref{eq:rwall} \citep{muzerolle03}.

\subsection{Inclinations} \label{sec:inclinations}
We infer the inclinations of our targets by estimating their stellar rotation periods from their \tess\ light curves and taking measurements of \vsini\ from \cite{manara21} and \cite{kounkel19}. Temporal analysis of the light curves uses the Scargle
periodogram analysis \citep{scargle82,horne86} as implemented in IDL\footnote{J\"orn Wilms 2005: \url{http://astro.uni-tuebingen.de/software/idl/aitlib/timing/scargle.html}}. 
We look for peaks in the power spectral density (PSD) between 1 to 10 days
in excess of a 99\% confidence level.

When power is found, we fold the data on the periods that show significant
power. To minimize long-term secular variability, we construct a running mean 
of width 1.5 times the period, and subtract that from the light curve prior
to folding. We bin the data into 20 phase bins, setting the uncertainty in
each phase bin to the variance in that bin and test the binned light curve
against the null hypothesis.

We also examine the auto-correlation of the light curve. The width of the
correlation peak is proportional to the duration of the typical features
contributing to the variations. We also consider the positive and negative
excursions in the light curve separately: positive excursions may be due
to bright patches on the photosphere; negative excursions may be due to
occultations of the surface by circumstellar material, or by starspots.
To define positive and negative excursions, we de-trend the light curve with
a polynomial fit (3$^{\rm rd}$ to 6$^{\rm th}$ order, depending on the
number of points) and retain only those points above or below the trend. Brief discussions of the individual targets can be found in Appendix~\ref{sec:obj_comments}.

\section{Analysis and Results} \label{sec:results}
\begin{figure*}[h!]
    \centering
    \plotone{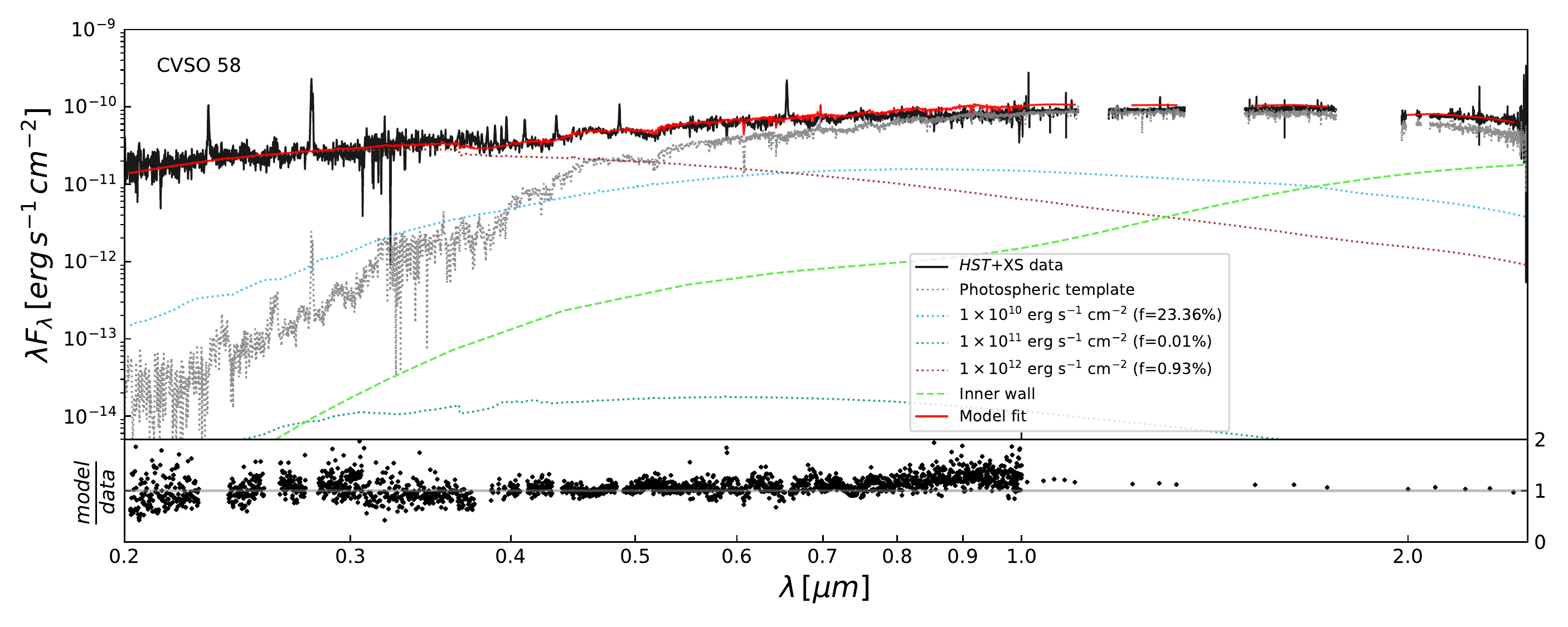}
    \plotone{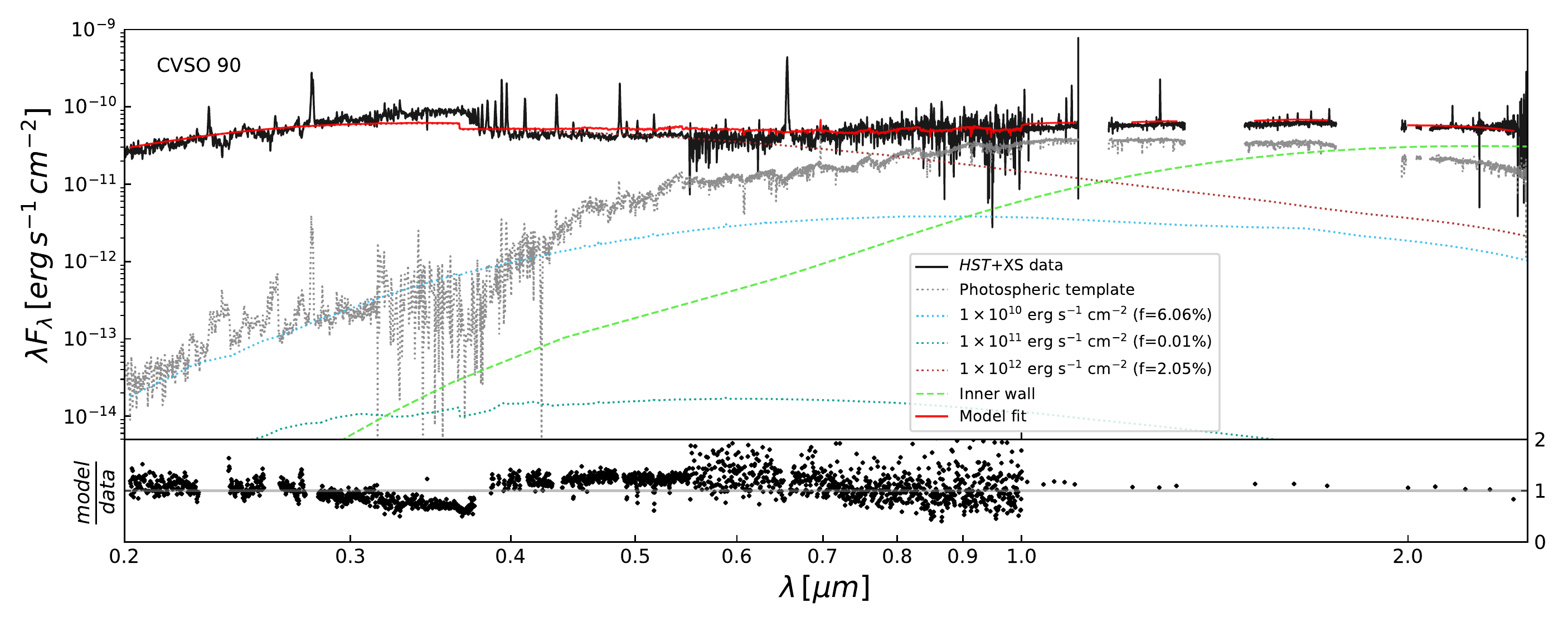}
    \plotone{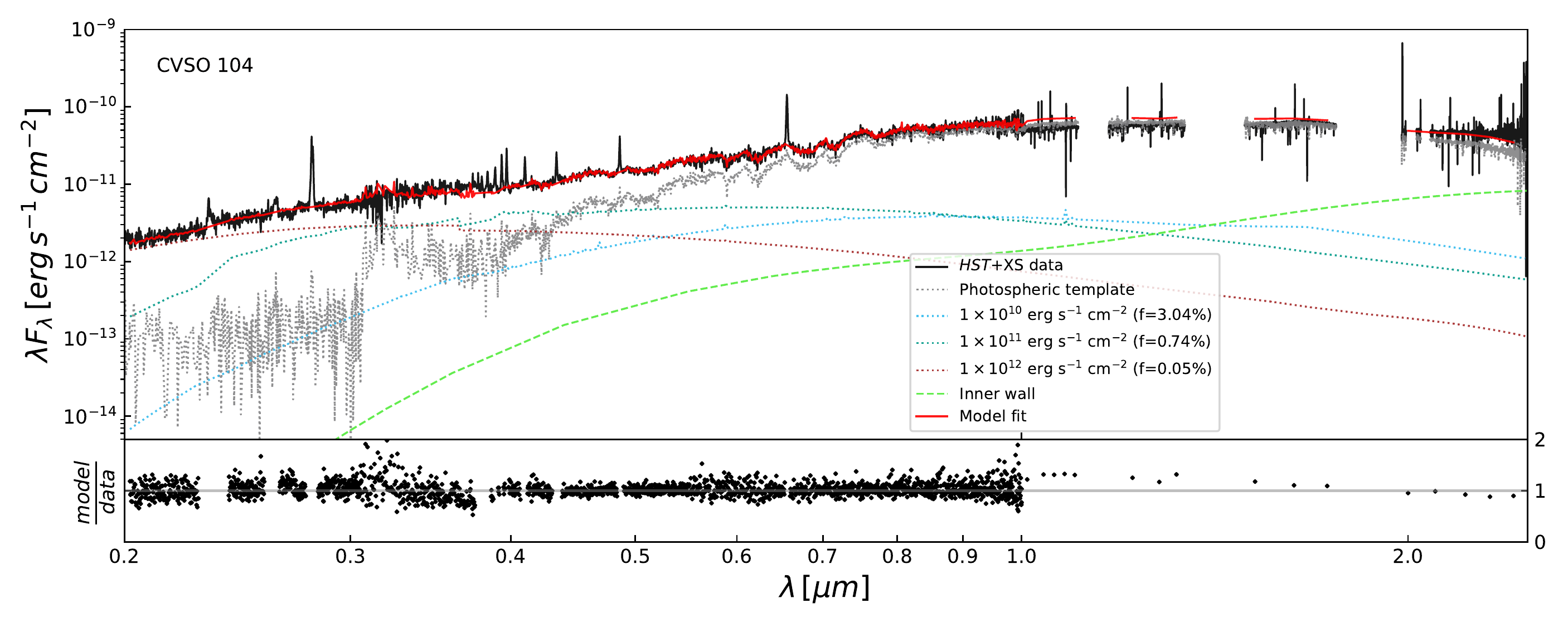}
    \caption{{Fitting the NUV--NIR continua of Orion OB1b targets. We show the {\it HST} (0.2--1~\micron) and X-Shooter (1--2.4~\micron) data (black) along with the photospheric template (grey), which is scaled using the optical veiling derived from VLT spectra listed in Table~\ref{tab:stellar_params}. The single model with the highest likelihood is shown in red. The median model parameter values and their associated 1$\sigma$ uncertainties are listed in Tables~\ref{tab:shockmodelparams} and \ref{tab:diskmodelparams}. The best fit model is a combination of the accretion shock model (consisting of the three columns with energy fluxes of 1$\times10^{10}$, 1$\times10^{11}$, and 1$\times10^{12}$ \escm|cyan, sea-green, and brown dotted lines, respectively|with filling factors as indicated in the legend) and the DIAD model (consisting of the emission from the inner disk wall; lime green dashed line).} Residuals are shown in the bottom panel of each figure. Emission lines were excluded from the fit, and regions significantly affected by telluric absorption were excluded from both the observed spectra and model fitting; this produces the gaps seen in the residual plots.}
    \label{fig:model_fits}
\end{figure*}
\begin{figure*}[h!]
    \centering
    \plotone{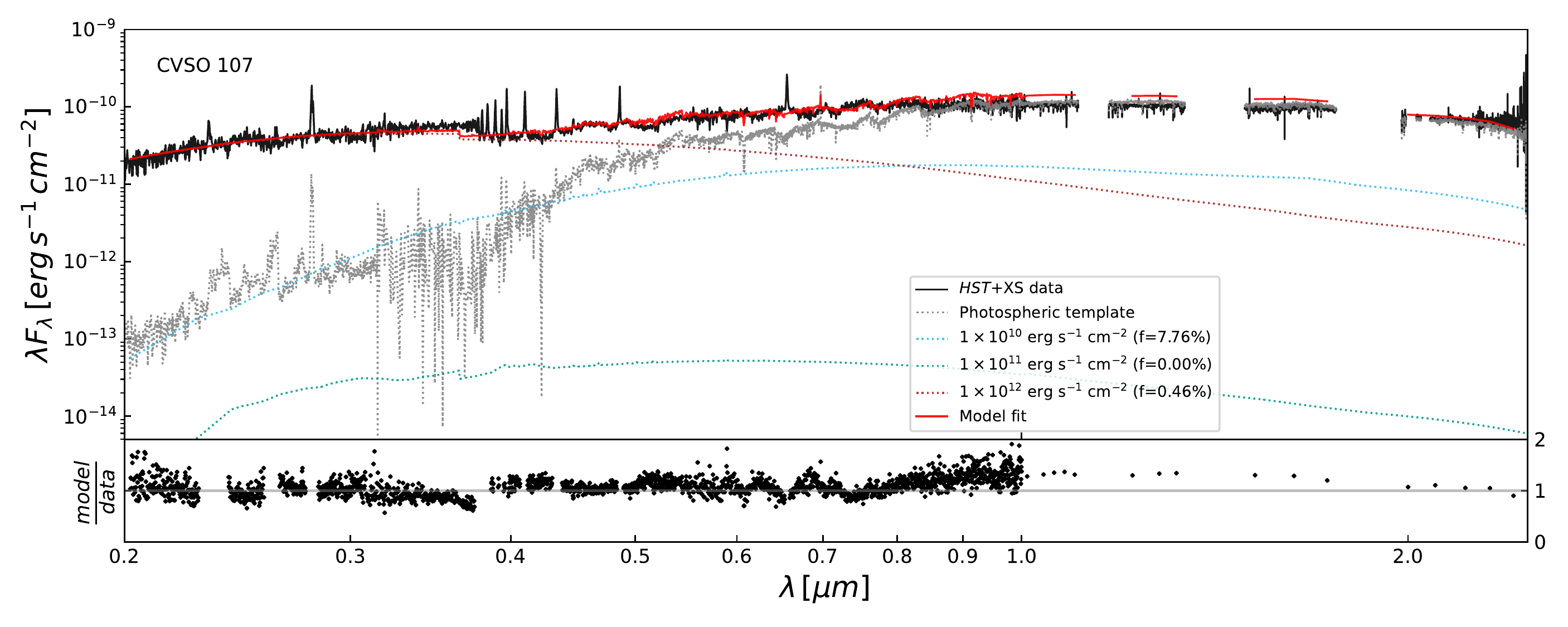}
    \plotone{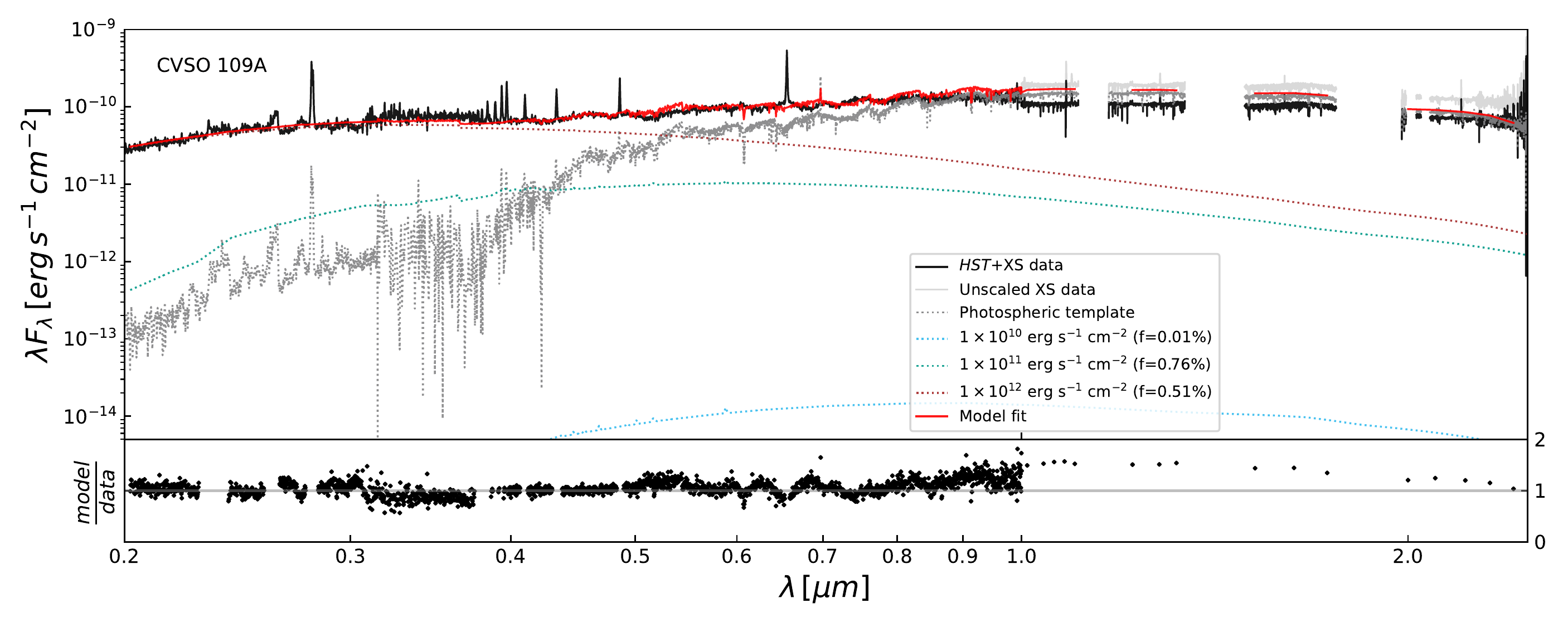}
    \plotone{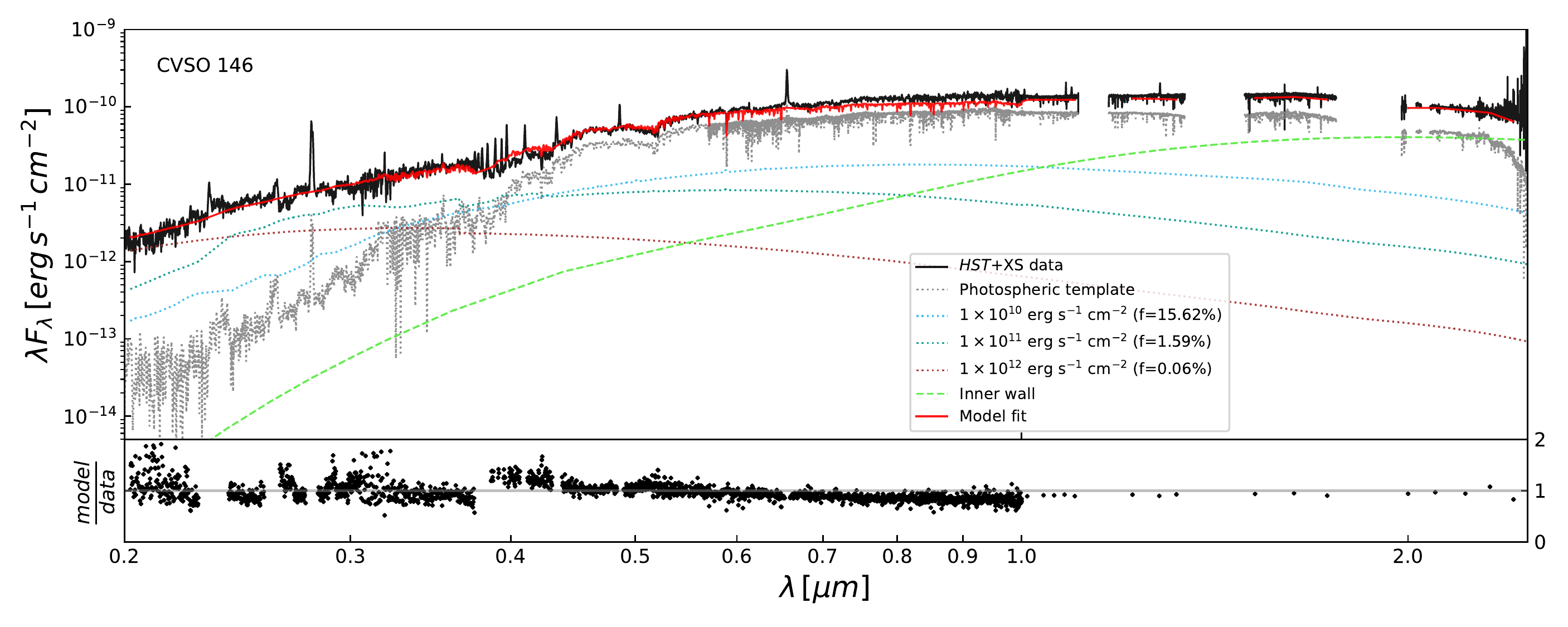}
    \tablecomments{Figure \ref{fig:model_fits} continued. The unscaled X-Shooter spectrum of CVSO~109 is shown in light grey.}
\end{figure*}
\begin{figure*}[h!]
    \plotone{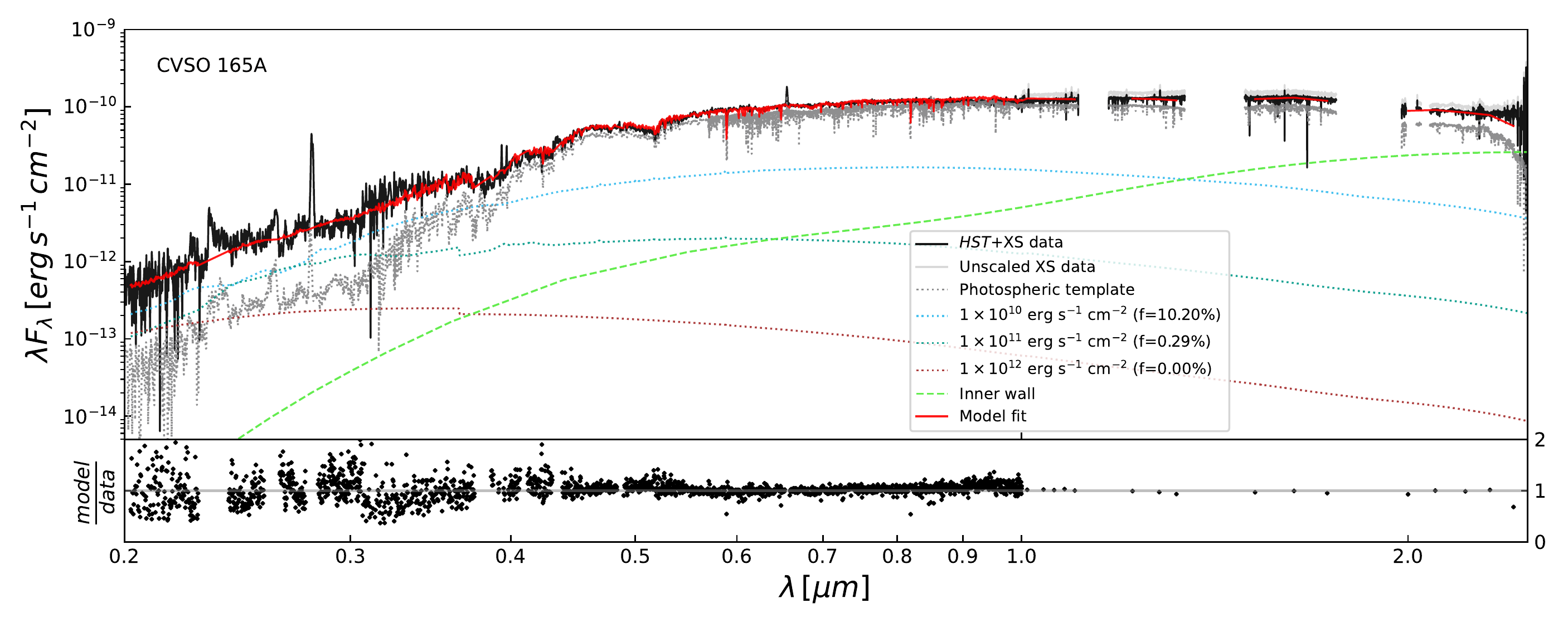}
    \plotone{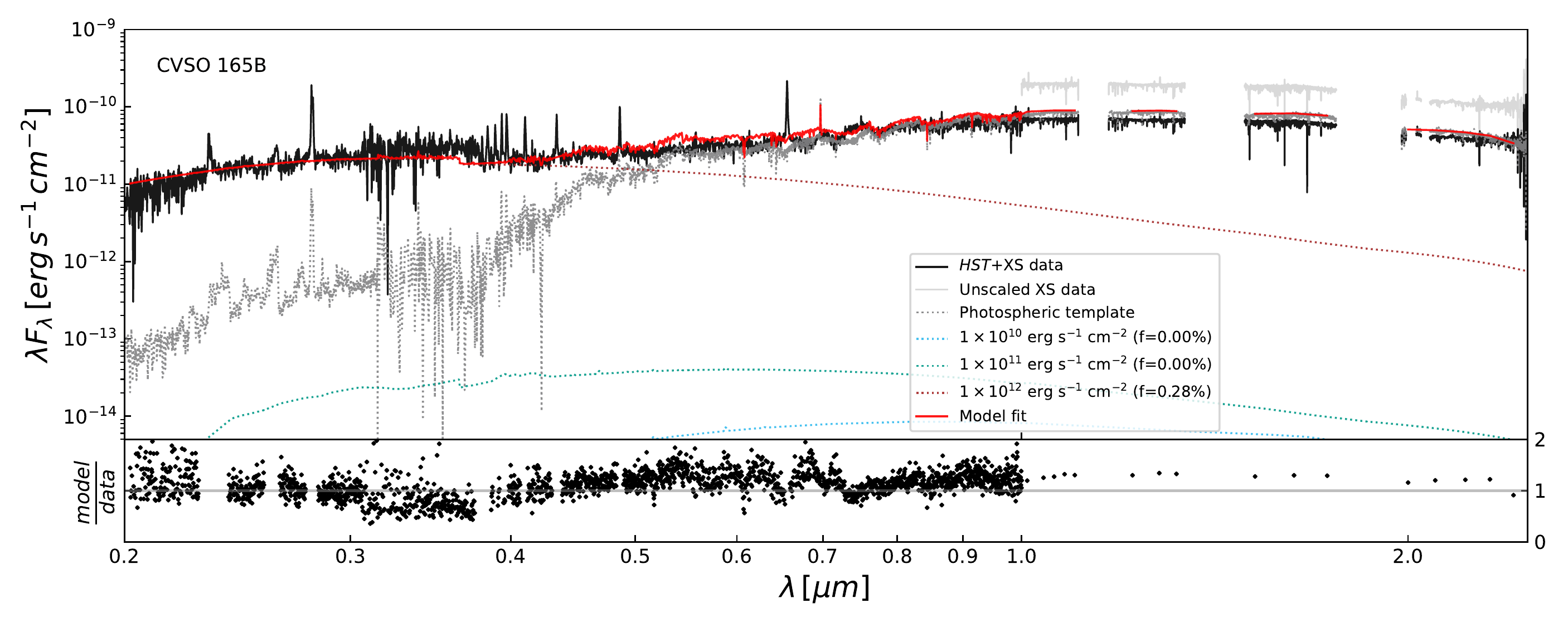}
    \plotone{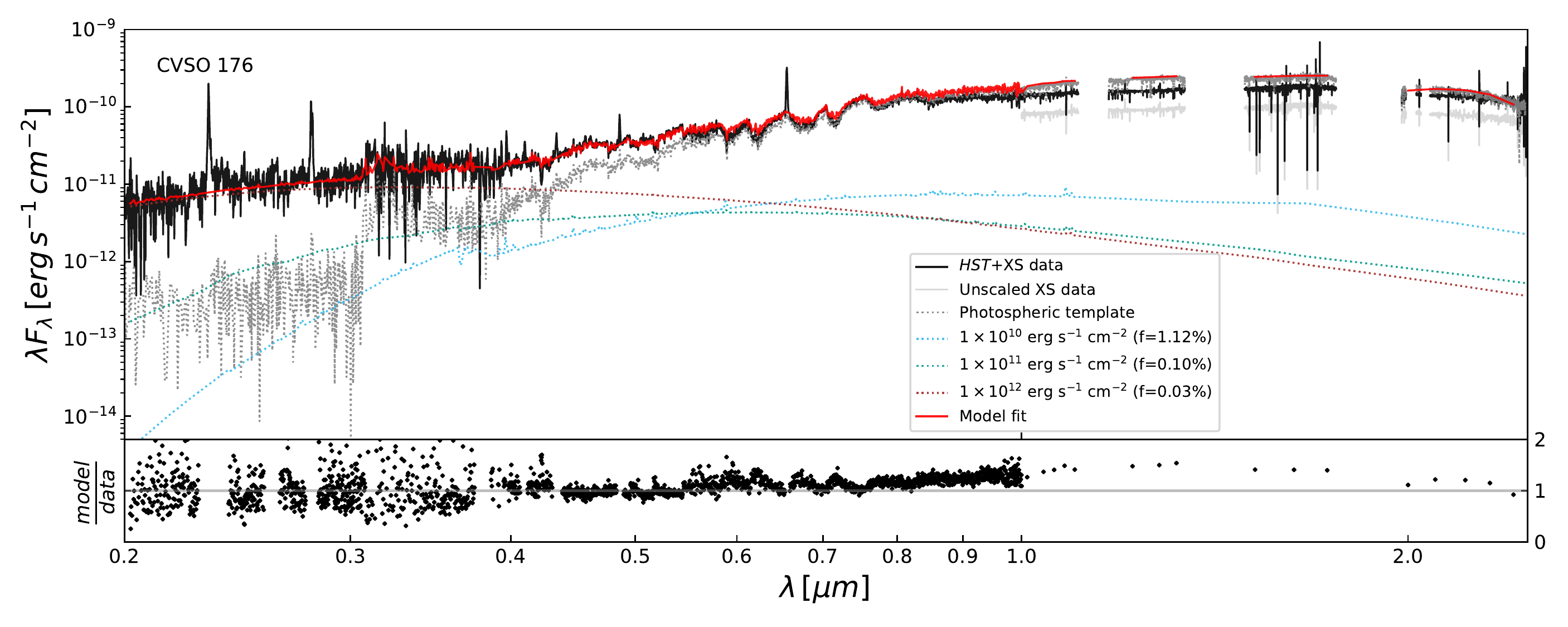}
    \tablecomments{Figure \ref{fig:model_fits} continued. The X-Shooter data for CVSO~165 are unresolved, so the NIR spectra have been scaled to align with the \hst\ continuum of each component. This is just an approximation given that the two components have different spectral types. The X-Shooter spectrum of CVSO~176 has been scaled up to the \hst\ continuum by a factor of 1.75 to account for the discontinuity between the continua, which is likely due to variability given that the observations were $\sim$65 hours apart. The unscaled X-Shooter spectra are shown in light grey.}
\end{figure*}

Figure~\ref{fig:model_fits} shows the best fits of the accretion shock and accretion disk models to the full NUV--NIR spectra of the nine CTTS analyzed here. Tables~\ref{tab:shockmodelparams} and \ref{tab:diskmodelparams} list, respectively, the accretion and disk model parameters associated with each fit. Using our accretion shock model, we derive V-band extinction, stellar radius, accretion rate, and accretion column structure for each target.

Our results indicate that accretion remains strong in TTS longer than originally expected. \citet{hartmann16} predict an accretion rate on the order of $3\times 10^{-9}$ \msunyr\ for a 0.7~\msun\ star of 5~Myr (the median age of the Orion OB1b region). However, \citet{ingleby14} measured accretion rates on the order of $1\times 10^{-8}$ \msunyr\ for CVSO~58, CVSO~90, and CVSO~109. Similarly, our derived accretion rates range from 0.5--17.2$\times10^{-8}$ \msunyr, with a median value of 1.2$\times10^{-8}$ \msunyr. These accretion rates are comparable to those of the 1--2~Myr regions Taurus and Chamaeleon I \citep{ingleby13}. These high accretion rates produce associated accretion luminosities that range from 0.07--1.96~$L_\star$, with a median accretion luminosity of 0.25~$L_\star$. Given that the accretion luminosities are comparable to the stellar luminosities for these targets, the accretion rates from the accretion shock model are important inputs to the inner disk wall model.

The average accretion shock temperatures range from 4800~K to 10542~K, in agreement with the range predicted by \cite{cg98}. With a mean temperature of 6511~K, these targets have $T_{\rm shock}$ notably lower than the typically-assumed temperature of 8000--10000~K for a single accretion column \citep[e.g.,][]{dalessio98,fischer11}. The average temperatures of the three accretion columns are 4274~K for the low flux-density column; 6695~K for the medium flux-density column; and 10786~K for the high flux-density column. See Table~\ref{tab:Tshock} in the appendix for the best-fit temperatures for each column of each target.

We are able to attain satisfactory fits to the NIR excesses of five targets, though there are a number of cases in which the photosphere goes above the data around 1~\micron\ and thus forces the total model fit to be above the data (CVSO~58, CVSO~104, CVSO~107, CVSO~109A, CVSO~165B, and CVSO~176). In the cases of CVSO~107, CVSO~109A, CVSO~165B, and CVSO~176, there is no clear NIR excess above the photospheric and accretion shock emission, so no inner disk wall is fitted to these targets. This may result from variability in the veiling, which dictates the scaling of the WTTS photospheric template. This will be examined further in future work. 

The lack of NUV spectra of WTTS templates with the same spectral subtype as some of the CTTS may produce scaling that results in mismatches in the NIR. Four out of the five targets that are fit with an inner wall require a wall height in excess of 5 gas scale heights to produce enough emission to account for the excess. These targets are likely pre-transitional disks, which have extra emission from an optically-thin dust component in their inner disk cavities \citep{mauco18}. Our future work will examine whether this additional component produces a better fit to the data.

\subsection{Uncertainties}
The average accretion shock model uncertainty, quantified by an MCMC nuisance parameter, is 14\%. This statistical goodness-of-fit is a lower limit to the actual uncertainty as it does not take the uncertainties of the input parameters, the photospheric templates, or the extinction curve into account. When accretion rates are calculated considering the uncertainties in \rstar\ and the filling factors, and assuming an uncertainty of 2\rstar\ for the magnetospheric truncation radius \ri, the median percent error increases from 4\% to 32\%.

Uncertainties in our derived stellar radii take into account the error in the data fluxes, CTTS distance, veiling, visual extinction, and an assumed 10\% error in the WTTS template fluxes. Our estimates have an average percent error of 8\%. The updated luminosity measurements included in Table~\ref{tab:shockmodelparams} propagate our derived error in \rstar, but no uncertainty is assumed for the stellar temperature.
The uncertainties in the average shock temperature are dominated by the errors in the filling factors. Since the errors are often asymmetric, the larger of the upper and lower uncertainties is taken as the error in propagation.

A significant source of systematic uncertainty comes from the availability and scaling of the WTTS photospheric templates. There are seven total WTTS templates with NUV spectra available between SpTs K5 and M2, and there are no templates of type K6 or M0. All CTTS studied here are fit with a template within 1.5 subtypes of their own spectral classification, but even this difference in spectral type introduces some amount of error. Fitting each target CTTS with the next closest template on either side, when available, changes the best-fit visual extinction by 0.37 mag on average.

Once a template is chosen, it must be scaled to the data by the veiling measurement, which has its own variability and uncertainty. Three measurements of veiling, each separated by a day, were available for all targets except CVSO~90 and CVSO~104. The measured veiling for a given target varied, on average, by a factor of 1.6 across the three epochs. By choosing the epoch closest in time to the \hst\ observations, we minimize the influence of veiling variability, but it cannot be completely removed.

Our derivation of inclinations are rough estimates given that the stellar rotation periods in the \tess\ data can easily be obscured by other processes, such as occultation dips and discrete accretion events. Uncertainties have not been determined for the period measurements, so the quoted uncertainties in inclinations come from errors in \vsini\ and \rstar. We note that if the inclination estimates are wrong, the only parameter that would be affected is the height of the inner disk wall $z_{\rm wall}$.

\begin{deluxetable}{lccccccccc}
\rotate
\tablecaption{Best-fit Accretion Shock Model Parameters of Orion OB1 CTTS
\label{tab:shockmodelparams}} 
\tablehead{
 \colhead{Object}  & \colhead{$A_{V}$}  & \colhead{$R_{\star}$} & \colhead{$L_{\star}$} & \colhead{$L_{\rm acc}$} & \colhead{$\dot{M}$} & \colhead{$f_{1E10}$} & \colhead{$f_{1E11}$} & \colhead{$f_{1E12}$} & \colhead{$T_{\rm shock}$} \\
 \multicolumn{1}{c}{}  & \multicolumn{1}{c}{(mag)} & \multicolumn{1}{c}{(\rsun)} & \multicolumn{1}{c}{(\lsun)} & \multicolumn{1}{c}{(\lsun)} & \multicolumn{1}{c}{(10$^{-8}$ \msunyr)} & \multicolumn{3}{c}{(fraction of stellar surface covered)} & \multicolumn{1}{c}{(K)}
}
\startdata
CVSO~58 & $1.39^{+0.04}_{-0.04}$ & $1.05^{+0.07}_{-0.07}$ & 0.25$\pm$0.03 & $0.202^{+0.011}_{-0.011}$ & $1.03^{+0.05}_{-0.06}$ & $0.231^{+0.010}_{-0.022}$ & $0.00020^{+0.0017}_{-0.00013}$ & $0.0092^{+0.0007}_{-0.0007}$ & 5493$\pm$96 \\
CVSO~90 & $0.92^{+0.03}_{-0.30}$ & $0.84^{+0.09}_{-0.09}$ & 0.12$\pm$0.03 & $0.234^{+0.011}_{-0.08}$ & $1.25^{+0.06}_{-0.4}$ & $0.053^{+0.016}_{-0.03}$ & $0.00031^{+0.024}_{-0.00023}$ & $0.0202^{+0.0011}_{-0.010}$ & 7975$\pm$1105 \\
CVSO~104 & $0.05^{+0.05}_{-0.03}$ & $1.64^{+0.13}_{-0.12}$ & 0.36$\pm$0.06 & $0.0655^{+0.003}_{-0.0021}$ & $1.14^{+0.05}_{-0.04}$ & $0.0304^{+0.0027}_{-0.004}$ & $0.0074^{+0.0005}_{-0.0004}$ & $0.00049^{+0.00006}_{-0.00003}$ & 5239$\pm$105 \\
CVSO~107 & $1.16^{+0.03}_{-0.05}$ & $1.97^{+0.15}_{-0.15}$ & 0.66$\pm$0.10 & $0.331^{+0.012}_{-0.021}$ & $4.84^{+0.17}_{-0.3}$ & $0.076^{+0.004}_{-0.007}$ & $0.00015^{+0.0027}_{-0.00008}$ & $0.00456^{+0.00024}_{-0.0004}$ & 5656$\pm$128 \\
CVSO~109A & $0.83^{+0.04}_{-0.32}$ & $2.55^{+0.20}_{-0.20}$ & 1.18$\pm$0.19 & $0.61^{+0.04}_{-0.23}$ & $17.2^{+1.0}_{-7}$ & $0.0006^{+0.011}_{-0.0005}$ & $0.0073^{+0.0026}_{-0.0021}$ & $0.0051^{+0.0005}_{-0.0026}$ & 8877$\pm$1910 \\
CVSO~146 & $0.28^{+0.03}_{-0.02}$ & $1.25^{+0.09}_{-0.09}$ & 0.37$\pm$0.05 & $0.0926^{+0.003}_{-0.0027}$ & $0.530^{+0.017}_{-0.015}$ & $0.155^{+0.005}_{-0.006}$ & $0.0159^{+0.0011}_{-0.0010}$ & $0.00058^{+0.00005}_{-0.00004}$ & 4975$\pm$27 \\
CVSO 165A & $0.33^{+0.02}_{-0.02}$ & $1.69^{+0.20}_{-0.20}$ & 0.56$\pm$0.13 & $0.0617^{+0.0017}_{-0.0015}$ & $0.708^{+0.020}_{-0.017}$ & $0.1020^{+0.0024}_{-0.003}$ & $0.00287^{+0.0004}_{-0.00028}$ & $0.0000485^{+0.000007}_{-0.0000023}$ & 4800$\pm$14 \\
CVSO 165B & $1.23^{+0.02}_{-0.02}$ & $2.00^{+0.25}_{-0.25}$ & 1.14$\pm$0.29 & $0.176^{+0.005}_{-0.006}$ & $1.65^{+0.05}_{-0.06}$ & $0.00013^{+0.0004}_{-0.00007}$ & $0.00009^{+0.00015}_{-0.00003}$ & $0.00275^{+0.00008}_{-0.00012}$ & 10542$\pm$343 \\
CVSO~176 & $1.44^{+0.03}_{-0.03}$ & $3.26^{+0.31}_{-0.31}$ & 1.08$\pm$0.21 & $0.0815^{+0.0023}_{-0.0020}$ & $4.18^{+0.12}_{-0.10}$ & $0.0108^{+0.0019}_{-0.004}$ & $0.00106^{+0.00026}_{-0.00020}$ & $0.000267^{+0.000017}_{-0.000015}$ & 5043$\pm$254 \\
\enddata
\tablecomments{Best-fit accretion shock model parameters for each CTTS. To calculate the models, the stellar mass, distance, temperature, and veiling are adopted as given in Table~\ref{tab:stellar_params}. \rin\ is taken to be 5\rstar\ for all targets except CVSO~109A, which has $R_{\rm i}=2.3$\rstar\ as described in Section~\ref{sec:shockmodel}. For the best-fit temperatures of individual accretion columns, see Table~\ref{tab:Tshock} in the appendix.}
\end{deluxetable}

\begin{deluxetable}{lcccc}
\tabletypesize{\footnotesize}
\tablecaption{Periods and Inclinations of Orion OB1 CTTS
\label{tab:inclinations}}
\tablehead{
 \colhead{Object} & \colhead{\vsini} & \colhead{Period} & \colhead{$v$} & \colhead{$i$} \\
 \multicolumn{1}{c}{} & \multicolumn{1}{c}{(km/s)} & \multicolumn{1}{c}{(days)} & \multicolumn{1}{c}{(km/s)} & \multicolumn{1}{c}{($^{\circ}$)}
}
\startdata
CVSO~58 & 17.9$\pm$1.3 & 5.7 & 9.3$\pm$0.5 & ... \\
CVSO~90 & 8.3$\pm$1.6 & 5.1 & 8.3$\pm$0.9 & ... \\
CVSO~104 & 7.5$\pm$1.0 & 4.7 & 17.7$\pm$1.3 & 25.1$\pm$4.1 \\
CVSO~107 & 5.9$\pm$0.9 & 6.4 & 15.6$\pm$0.9 & 22.3$\pm$3.9 \\
CVSO~109A & 3.2$\pm$0.9 & 6.5 & 19.8$\pm$1.2 & 9.3$\pm$2.7 \\
CVSO~146 & 5.0$\pm$0.8 & 5.5 & 11.5$\pm$0.7 & 25.8$\pm$4.8 \\
CVSO~165A & 15.4$\pm$0.9 & 4.3 & 19.9$\pm$1.4 & 50.8$\pm$6.5 \\
CVSO~165B & 15.4$\pm$0.9 & 4.3 & 23.5$\pm$1.9 & 40.9$\pm$4.9 \\
CVSO~176 & 18.4$\pm$1.2 & 7.1 & 23.2$\pm$1.9 & 52.4$\pm$7.8 \\ 
\enddata
\tablecomments{Calculated inclinations $i$ for our targets. \vsini\ comes from the VLT modeling that produced the veilings used in this paper \citep{manara21} for all targets except CVSO~90, which takes \vsini\ from \cite{kounkel19}. Stellar rotational velocities $v$ are calculated using the radii we derive here. The derived $v$ for CVSO~58 is smaller than its \vsini, and the calculated $i$ for CVSO~90 has an error of almost $\pm180^\circ$; thus, inclination is assumed to be a standard 60$^\circ$ for these targets, corresponding to cos($i)=0.5$.}
\end{deluxetable}

\begin{deluxetable}{lccc}
\tablecaption{Best-fit Accretion Disk Model Parameters of Orion OB1 CTTS
\label{tab:diskmodelparams}}
\tablehead{
 \colhead{Object} & \colhead{$z_{\rm wall}$} & \colhead{$T_{\rm wall}$} & \colhead{$R_{\rm wall}$} \\
 \multicolumn{1}{c}{} & \multicolumn{1}{c}{(H)} & \multicolumn{1}{c}{(K)} & \multicolumn{1}{c}{(AU)}
}
\startdata
CVSO~58 & 5 & 1200 & 0.10 \\ 
CVSO~90 & 5.5 & 1600 & 0.06 \\ 
CVSO~104 & 5.5 & 1200 & 0.07 \\ 
CVSO~146 & 18 & 1800 & 0.05 \\ 
CVSO~165A & 8 & 1400 & 0.07 \\ 
\enddata
\tablecomments{Best-fit parameters for the the accretion disk models given input parameters as shown in Tables~\ref{tab:stellar_params} and \ref{tab:shockmodelparams}. We adopt $a_{\rm min}=0.005$~{\micron}, $a_{\rm max}=10$~{\micron}, and $p=3.5$. Disk inclinations $i$ are taken from Table~\ref{tab:inclinations}.}
\end{deluxetable} 

\section{Discussion}
\label{sec:discussion}
In Section~\ref{sec:lit_comp}, we place our results in the context of previous studies. In Section~\ref{sec:Av}, we examine the significance of the choice of extinction law and the wavelength range of data available.

\subsection{Comparison to previous results}
\label{sec:lit_comp}
\begin{deluxetable*}{lccccccccccc}
\tablecaption{Literature comparison  \label{tab:lit_comp}}
\tablehead{
\colhead{Object} & \colhead{$\dot{M}_{\rm C05}$} & \colhead{\av$_{\rm ,C05}$}  & \colhead{$\dot{M}_{\rm I14}$} & \colhead{\av$_{\rm ,I14}$}  & \colhead{$\dot{M}_{\rm M18}$} & \colhead{\av$_{\rm ,M18}$}  & \colhead{$\dot{M}_{\rm M21}$} & \colhead{\av$_{\rm ,M21}$} & \colhead{BC$^*$}  & \colhead{$\dot{M}_{\rm P22}$} & \colhead{\av$_{\rm ,P22}$} \\
\cmidrule(lr){2-9} \cmidrule(lr){10-12}
\colhead{} & \colhead{($\times10^{-8}$)} & \colhead{(mag)} & \colhead{($\times10^{-8}$)} & \colhead{(mag)}& \colhead{($\times10^{-8}$)} & \colhead{(mag)}& \colhead{($\times10^{-8}$)} & \colhead{(mag)} & \colhead{} &  \colhead{($\times10^{-8}$)} & \colhead{(mag)}
}
\startdata
CVSO~58 & 0.45 & 0.12 & 1.60 & 0.8$\pm0.5$ & … & … & 0.43 & 0.8 & &  $1.03^{+0.06}_{-0.06}$ & $1.39^{+0.04}_{-0.04}$ \\
CVSO~90 & 1.77 & 0.00 & 1.00 & 0.0$\pm0.4$ & … & … & 0.25 & 0.1 & &  $1.25^{+0.05}_{-0.5}$ & $0.92^{+0.03}_{-0.30}$ \\
CVSO~104 & 0.75 & 0.00 & … & … & 0.56 & 0.1 & 0.32 & 0.2 & &  $1.14^{+0.06}_{-0.04}$ & $0.05^{+0.05}_{-0.03}$ \\
CVSO~107 & 1.09 & 0.32 & 0.25 & 0.7$\pm0.4$ & 0.29 & 0.4 & 5.01 & 0.3 & &  $4.83^{+0.19}_{-0.29}$ & $1.16^{+0.03}_{-0.05}$ \\
CVSO~109 & 2.52 & 0.00 & 3.00 & 0.8$\pm0.5$ & 0.67 & 0.0 & 3.24 & 0.1 & A &  $17.2^{+1.0}_{-6}$ & $0.83^{+0.04}_{-0.32}$ \\
CVSO~146 & 0.81 & 0.37 & … & … & … & … & 0.27 & 0.6 & &  $0.529^{+0.016}_{-0.015}$ & $0.28^{+0.03}_{-0.02}$ \\
CVSO~165 & 0.37 & 0.00 & … & … & … & … & 0.08 & 0.2 & A &  $0.708^{+0.021}_{-0.017}$ & $0.33^{+0.02}_{-0.02}$ \\
 & … & … & … & … & … & … & … & … & B & $1.65^{+0.05}_{-0.06}$ & $1.23^{+0.02}_{-0.02}$ \\
CVSO~176 & 0.43 & 0.00 & … & … & … & … & 1.45 & 1 & & $4.19^{+0.12}_{-0.11}$ & $1.44^{+0.03}_{-0.03}$ \\
\enddata
\tablecomments{Comparison between the accretion rates and visual extinctions derived here and those derived by \cite{calvet05} (C05), \cite{ingleby14} (I14), \cite{mauco18} (M18), and \cite{manara21} (M21).
$^*$For the binary targets that are resolved in this study, the BC column indicates which binary component is being referenced.}
\end{deluxetable*}

\cite{re19} applied the multi-column accretion shock model from \cite{ingleby13} to multi-epoch observations of five CTTS and found that mass accretion rates vary with a spread of a factor of $\sim$2 for a given object. Similarly, \citet{venuti14} found that the accretion rate for a given object has a spread of 0.5 dex (a factor of $\sim$3) from studying about 200 CTTS in NGC 2264. Rather than merely representing this intrinsic variability, the best-fit values for accretion rate and visual extinction presented in this work are systematically higher than those found by previous studies of objects in this sample \citep{calvet05,ingleby14,mauco18,manara21}. See Table~\ref{tab:lit_comp} for the individual values found by each work.

\citet{calvet05} modeled all CTTS analyzed here (though their observations of binaries CVSO~109 and CVSO~165 were unresolved), obtaining accretion rates for each using photometric excesses in U--B and U--V. They cite an overall uncertainty of a factor of 3 for each measured accretion rate, with the largest contribution coming from uncertainty in the extinction. They calculate \av\ from the $V-I_C$ color using the \citet{cardelli89} extinction law with $R_V=3.1$. The accretion rates measured here are 0.6-9.7 times that of those presented in \citet{calvet05}, with a median ratio of 2.3. This cannot be attributed to different flux levels, as all of our observed U--V colors are redder than those presented in \citet{calvet05}. Additionally, our measured extinctions are systematically larger than those of \citet{calvet05}, with a median difference of 0.8 mag. Because our accretion rates are systematically higher, it is likely that the different accretion rates result from systematic effects in the different modeling techniques and adopted extinctions rather than solely from true variability.

The closest analogue to our modeling technique is that of \citet{ingleby14}, who modeled CVSO~58, CVSO~90, CVSO~107, and the unresolved CVSO~109 system. They used the veiling at V and I to estimate extinction by comparing the observed photospheric V-I$_C$ colors to the standard colors in \citet{kh95}. They note that their inferred V-band extinctions would decrease by 0.2-0.4 mag had they used the colors of \citet{pm13}, and this in turn would decrease their \mdot s by a factor of 1.8–2.4. They then deredden their data using the \citet{whittet04} law towards HD 29647, as we do here. The MagE and MIKE spectra they used for these four targets covered only 3400--9000~\AA, so their fitting does not include the NUV.

\citet{ingleby14} thus fit the excess shock emission beyond the Balmer jump using a five-column accretion shock model also based on \citet{cg98} (the three columns used here plus two intermediate columns). However, instead of using the standard accretion shock model, they found that they needed to increase the low-density preshock emission by up to a factor of five to accurately reproduce the bluest regions of the excess emission spectra. This likely explains why the structure of their accretion columns, specifically the area they cover, is notably different from this work. Some of our model fits show a trend of underestimating the flux near the Balmer jump; future work will examine whether increasing this preshock emission produces a better fit to the data, or whether this is merely an effect of Balmer-line crowding that is not accounted for by our continuum model.

The ratios of our accretion rates to that of \cite{ingleby14} are 0.7 (CVSO~58), 1.2 (CVSO~90), 19.3 (CVSO~107), and 5.7 (CVSO~109). Note that had they used the same spectral type--color conversion as we did \citep{pm13}, these ratios would have been higher as described above. The differences in our derived accretion rates correspond to the differences in U- and V-band magnitudes between our respective observations. CVSO~58 was dimmer in both bands in our epoch of observations ($\Delta U=0.70$ mag, $\Delta V=0.38$ mag), and we find a lower accretion rate. The other three CTTS were brighter in both bands in our epoch | CVSO~90 ($\Delta U=-0.48$ mag, $\Delta V=-0.53$ mag), CVSO~107, ($\Delta U=-0.61$ mag, $\Delta V=-0.04$ mag), and CVSO~109A ($\Delta U=-1.11$ mag, $\Delta V=-0.09$ mag) | and we found higher accretion rates for these. The greatest increase in U-band belongs to CVSO~109A, which is consistent with the \tess\ light curve that shows that our \hst\ observation occurred near a local maximum in CVSO~109's light curve \citep{espaillat22}. These magnitudes come from observed, rather than dereddened, data, so differences in our treatments of extinction have no effect. Thus, our accretion rates are in broad agreement with those of \cite{ingleby14}.

\citet{mauco18} estimated accretion rates for CVSO~104, CVSO~107, and the unresolved CVSO~109 system using the \halpha-\mdot\ relation found for Taurus CTTS by \citet{ingleby13}: log(\mdot)=1.1($\pm0.3$)log(L$_{{\rm H}\alpha})-5.5(\pm0.8$), where the value of the \halpha\ line luminosity was estimated as its equivalent width times the continuum flux. \cite{espaillat22} found that for CVSO~109 this method produces a lower accretion rate than do the accretion shock models. This is what we find here, as the ratios of our accretion rates to theirs are 2.0 (CVSO~104), 16.7 (CVSO~107), and 25.7 (CVSO~109). The discrepancy may result from 1) the small number of sources used to determine this relation (10 CTTS), or 2) the assumption of a uniform distance for all Taurus CTTS (140 pc) in the determination of $L_{{\rm H}\alpha}$.

\citet{manara21} analyzed X-Shooter spectra of all of the ULLYSES Orion targets, using a hydrogen slab model to obtain their accretion properties. They fit for \av\ using the \cite{cardelli89} extinction law. X-Shooter spectra are available only beyond 3000~\AA, so the NUV is not included in their analysis. Our multi-column accretion shock model finds higher accretion rates for all targets except CVSO~107, which has the highest accretion rate in the analysis of \cite{manara21}. The median ratio of our accretion rates to theirs is 3.5.

Six of our CTTS targets have stellar radii from \cite{manara21} included in Table~\ref{tab:stellar_params}. There are no individual uncertainties available for the \cite{manara21} spectral types and luminosities, but previous work using their method of determining stellar radii showed that objects with spectral types earlier than M4.5 have percent errors less than 25\% \citep{alcala17}. Assuming a percentage error of 25\% for the radii calculated from the \cite{manara21} spectral types and luminosities, we find that our best-fit radii for CVSO~58, CVSO~90, and CVSO~104 are consistent within the errors; those fit for CVSO~107 and CVSO~176 are larger by a factor of 1.4 and 1.8, respectively; and that fit for CVSO~146 is lower by a factor of 1.5.

The systematically higher accretion rates presented in this work can in large part be attributed to a) our use of the accretion shock model as opposed to \halpha\ luminosity or an isothermal hydrogen slab model, and b) the larger NUV wavelength coverage of these data, as the NUV is the most important region for constraining the highest energy-flux density accretion column. The latter has the important implication that all ground-based measurements of CTTS accretion rates may be underestimated. Since \hst\ has a finite lifetime, future work should examine whether a correction factor can be determined to account for the systematic underestimation of accretion rates caused by the lack of NUV coverage.

\subsection{Extinction law}
\label{sec:Av}
When modeling NUV observations of CTTS, the choice of extinction law is incredibly important because of the strong attenuation by grains at UV wavelengths and the constraint imposed by the 2175~\AA\ bump. If a law produced for the general ISM is used, such as \citet{cardelli89} or \citet{fitzpatrick19}, the absorption bump feature is strong and will bias results towards lower values of \av. If, by contrast, the \cite{whittet04} law towards HD 29647 (which is embedded in Taurus) is used, the absorption feature is much less pronounced and the fitted \av\ will be higher.

No single interstellar extinction law can describe all star-forming regions equally well. This is supported by the finding that Taurus and Ophiuchus exhibit very different ultraviolet extinction functions, which \cite{whittet04} suggest is likely caused by their different populations of massive stars. Ophiuchus's significant population of OB stars produce radiation that maintains the strength of the 2175~\AA\ bump, so the \citet{cardelli89} law is better suited for use there than is the \cite{whittet04} law. Thus, NUV data for specific TTS provide an important constraint on the characteristics of both the individual stars themselves and their surrounding environments. Data from \hst\ are vital for probing the ISM of star-forming regions to distinguish between grain populations that significantly attenuate around 2175~\AA\ and those that do not.

\begin{figure}[h!]
    \centering
    \plotone{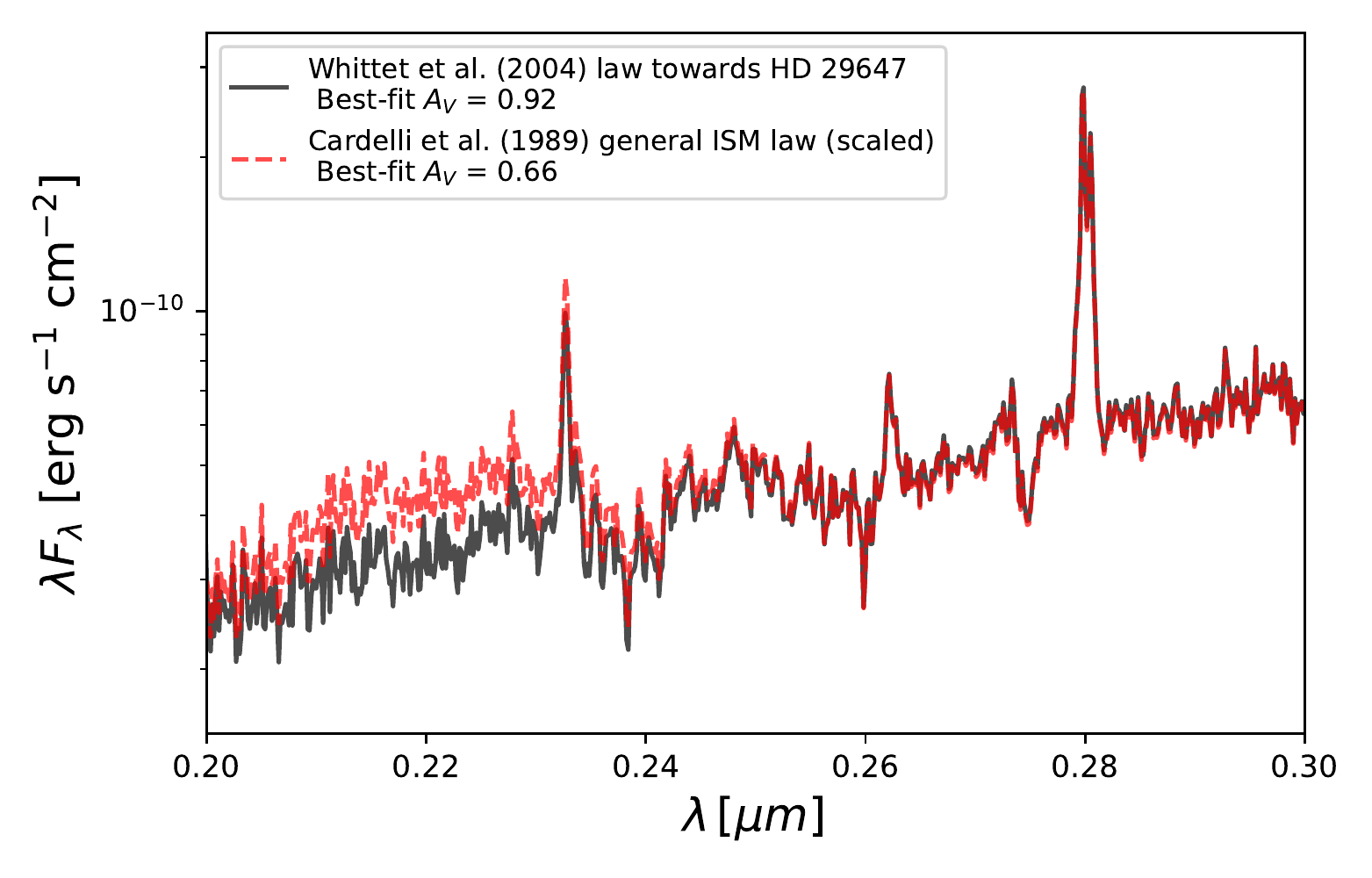} 
    \caption{Spectra of CVSO~90 dereddened with the best-fit V-band extinction when modeled with the \citet{whittet04} law (black solid line) and \citet{cardelli89} law (red dashed line, shown scaled to the \citet{whittet04} law spectrum at 3000~\AA). The overcorrected 2175~\AA\ bump is apparent in the red curve.}
    \label{fig:Av_comp}
\end{figure}

For these targets in Orion, the fits produced by the \citet{whittet04} law are better than those produced by the \citet{cardelli89} law, which overcorrects for the 2175~\AA\ bump (see Figure~\ref{fig:Av_comp}). This indicates that Orion OB1b is likely more similar to Taurus than Ophiuchus in terms of its interstellar extinction function. This is supported by an analysis of the ultraviolet interstellar radiation field (ISRF) of the region. The Habing field parameter $G_0$ gives the ratio of the local field enhanced by a neighboring OB star to the typical ISRF ($F_0$) according to

\begin{equation}
    G_0 = \frac{1}{F_0}\frac{L_{\rm FUV}}{4\pi r^2},
\end{equation}

\noindent where $F_0$ is assumed to be $1.6\times 10^{-3}$ \escm\ \citep{habing68}, $L_{\rm FUV}$ is approximated as the OB star's luminosity, and $r$ is the true distance between the OB star and the CTTS target of interest \citep{anderson13,mauco16}.

\cite{liseau99} found that the $\rho$ Ophiuchi star-forming region has $G_0=20$--140. In contrast, $G_0$ values for our OB1b CTTS targets are much lower. Assuming that the seven most significant OB stars around Orion OB1b ($\zeta$ Ori, $\epsilon$ Ori, $\delta$ Ori, $\eta$ Ori, 22 Ori, 25 Ori, and $\psi^2$ Ori) are all at the median OB1b distance of 400 pc, $G_0$ for our targets has a median value of 0.7 and a maximum value of 67. This, in combination with the low \av\ of the region, demonstrates that though Orion OB1b is an OB region, its interstellar extinction function should align more with that of quiescent Taurus than of hotter, dustier $\rho$ Ophiuchi. This can in part explain the relatively high accretion rates of this region in spite of its intermediate age, as these disks have not been externally photoevaporated by an enhanced ISRF. This is also consistent with the disk models from \cite{calvet05}, which could not reproduce the low long-wavelength infrared fluxes observed in Orion OB1 using small outer disk radii consistent with external photoevaporation ($\sim$30~AU). Instead, they concluded that the disks must be flatter and have larger maximum grain sizes than those in Taurus ($a_{\rm max}$=1~mm). For an analysis of the effect of an enhanced ISRF on circumstellar disk evolution, see \cite{anderson13}.

The ULLYSES survey, which will provide over 80 custom-calibrated NUV spectra of CTTS in nine star-forming regions, will allow us to examine the goodness-of-fit of different extinction laws in different environments. Beyond this, it would be ideal to attain extinction curves for all nearby star-forming regions rather than using either a general ISM curve or the curve calculated to the specific environment of Taurus.

\section{Summary}
\label{sec:summary}
\begin{itemize}
  \item Accretion rates for the 9 CTTS studied here range from  $0.5-17.2\times 10^{-8}$ \msunyr, relatively high for the intermediate age of Orion OB1b.
  \item Our accretion rates and V-band extinctions are systematically higher than those calculated from optical data in previous works, in large part due to our wavelength coverage that extends into the NUV.
  \item The NIR excesses of the five targets in which an excess is present are fit with 1200--1800~K inner disk walls located at 0.05--0.10~AU from the host stars.
  \item The choice of extinction law significantly affects the calculated accretion rate and introduces uncertainty that is difficult to quantify. Our analysis indicates that the environment of Orion OB1b is more similar to quiescent Taurus than to hot $\rho$ Ophiuchi. Ideally, extinction curves can be calculated for each star-formation region in the near future.
  \item This multi-column shock and DIAD analysis will be applied to all nine star-forming regions being observed through ULLYSES, allowing us to extend the analysis across distinct stellar populations and search for correlations between accretion and disk properties in a larger sample. Additionally, it will be applied to the four CTTS being monitored by the ULLYSES program (GM Aur, TW Hya, BP Tau, and RU Lup). 
\end{itemize}

\acknowledgements
{Support for this work comes from \hst\ AR-16129, as well as NASA through grant number AR 16129 from the Space Telescope Science Institute, which is operated by AURA, Inc., under NASA contract NAS 5-26555. This work benefited from discussions with the ODYSSEUS team (\url{https://sites.bu.edu/odysseus/}); see \cite{espaillat22} for an overview of the ODYSSEUS survey.

J.H. acknowledges support from CONACyT project No. 86372 and the UNAM-DGAPA-PAPIIT project IA102921. 
P.A., E.F. and \'A.K. acknowledge support from the European Research Council (ERC) under the European Union's Horizon 2020 research and innovation programme under grant agreement No 716155 (SACCRED). 
C.F.M. acknowledges funding by the European Union under the European Union’s Horizon Europe Research \& Innovation Programme 101039452 (WANDA). Views and opinions expressed are however those of the author(s) only and do not necessarily reflect those of the European Union or the European Research Council. Neither the European Union nor the granting authority can be held responsible for them.
C.F.M., J.M.A, M.G. and E.F. acknowledge support from the project PRIN-INAF 2019 ``Spectroscopically Tracing the Disk Dispersal Evolution''. S.H.P.A. acknowledges support from CNPq, CAPES
and Fapemig.

Based on observations obtained with the NASA/ESA Hubble Space Telescope, retrieved from the Mikulski Archive for Space Telescopes (MAST) at the Space Telescope Science Institute (STScI). STScI is operated by the Association of Universities for Research in Astronomy, Inc. under NASA contract NAS 5-26555.

This paper utilizes the D’Alessio irradiated accretion disk (DIAD) code. We wish to recognize the work of Paola D’Alessio, who passed away in 2013. Her legacy and pioneering work live on through her substantial contributions to the field.}

\bibliography{bibliography}{}
\bibliographystyle{aasjournal}

\appendix
\section{CVSO~17 and CVSO~36} \label{sec:app_A}
\begin{figure*}[h!]
    \centering
    \plottwo{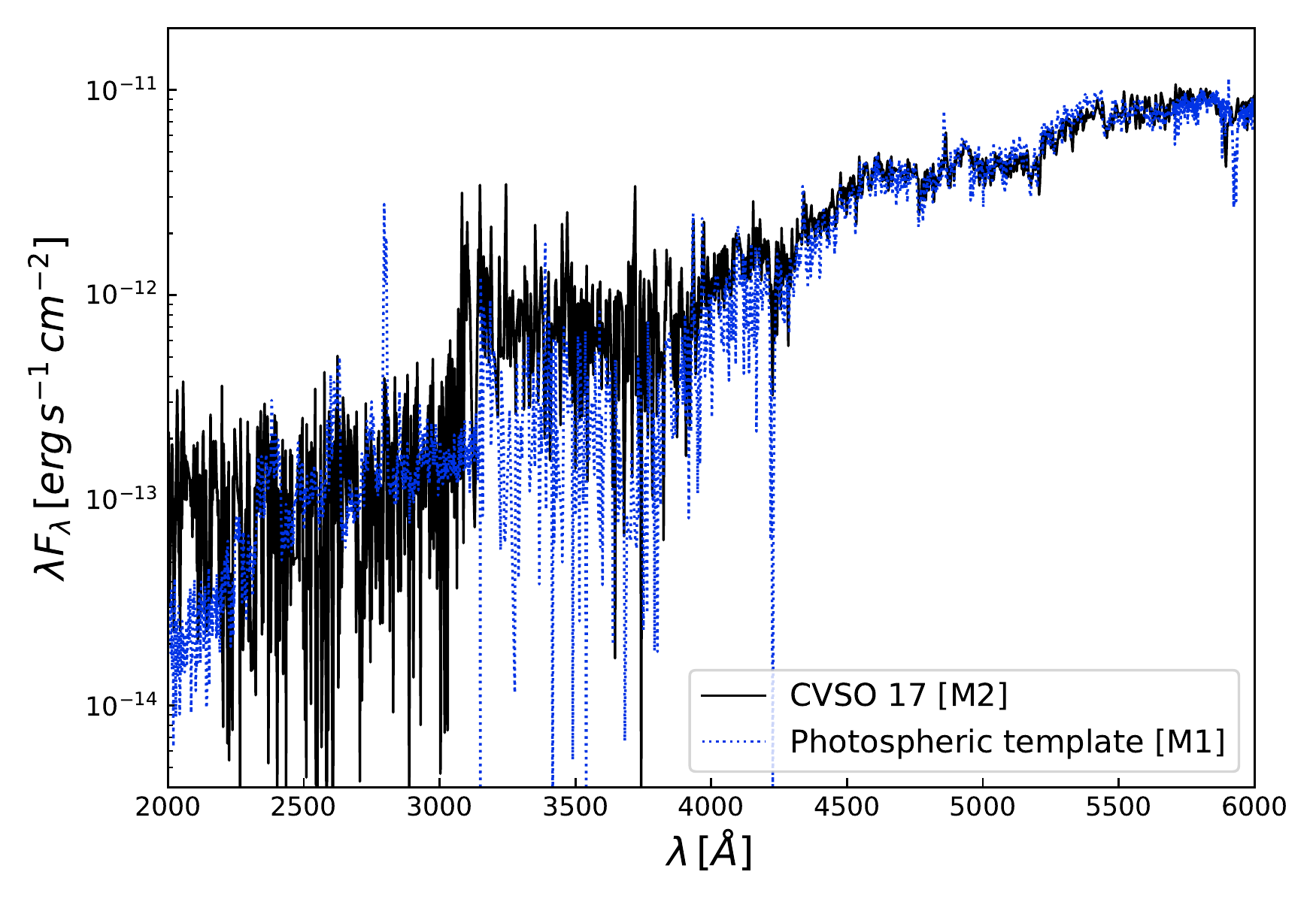}{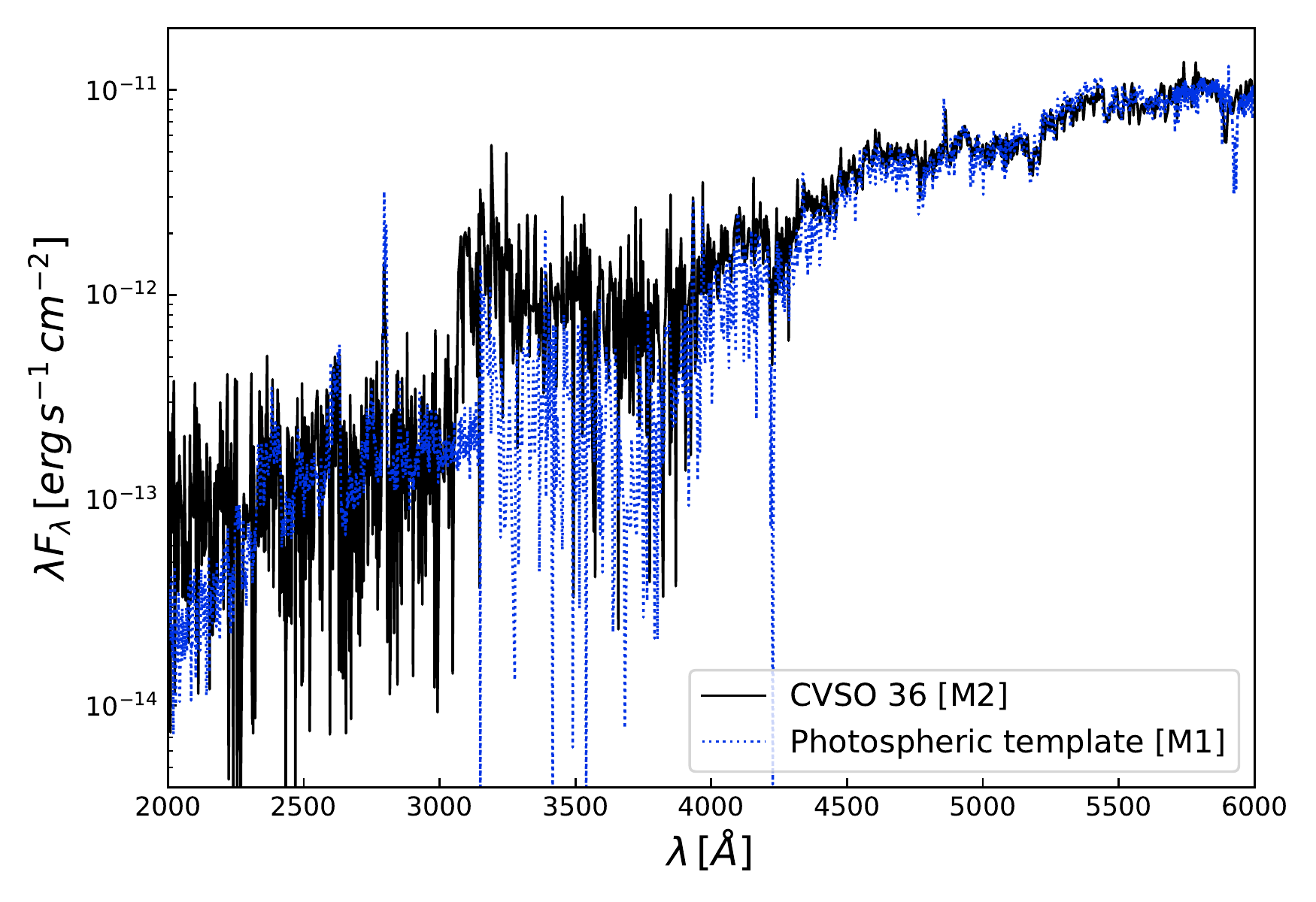}
    \caption{ The photospheric template shown is the M1 WTTS TWA 7.}
    \label{fig:17and36}
\end{figure*}

As shown in Figure~\ref{fig:17and36}, we find that CVSO~17 and CVSO~36 show no significant UV continuum excess in their \hst\ spectra that would indicate active accretion, which aligns with the analysis of their X-Shooter spectra by \cite{manara21} and confirms their photometric characterization as WTTS by \cite{calvet05}. Additionally, an analysis of their far-UV spectra shows no fluorescent H$_2$ emission, which is a clear delineation between CTTS and WTTS \citep{france12,alcala19}.

\section{TESS object notes} \label{sec:obj_comments}
{\bf CVSO~58:} 
In Sector~6 there was significant power at 2.87 and 5.31 days. 
The 2.9 day period folds well, but the 
likelihood that the amplitude of the period is significant is only 0.12.
In Sector~32 significant power exists at 6.45 and 1.41 days, but neither
looks periodic after folding.
The auto-correlation functions in both sectors show a peak near 5.7 days. 
This peak is strongest for the dips; during Sector~6 the positive excursions
(brightenings) show a less well-defined peak at a period near 6 days.

{\bf CVSO~90:} 
The Sector~6 PSD peaks at 3.15 days, but there is no significant period seen
in the folded light curve.
The Sector~32 PSD shows a strong and broad peak at 5.11 days. 
The folded light curve is not significant due to
lots of scatter in the half of the period dominated by deep dips. 
The autocorrelation function shows a peak at a lag of about 4.4 days for the
absorption dips, and a peak at a 5.7 day period for the brightenings.

{\bf CVSO~104:} 
This star has a peak in the PSD near 4.7 days in each sector.
It is not clear whether the light curve is dominated by bright or dark
excursions.
We do not attempt to fold the two sectors
together because it it impossible to keep track of the phases over this time.

{\bf CVSO~107:} 
The Sector~6 light curve is dominated by two deep dips spaced 12 days apart.
The periodogram finds power at 6.3 days, or half that spacing. Power at the 
same period also dominates in Sector~32. The brightenings show a preferred
lag of 6.5 days, while the dips lag at 7.7 days.
There are three possible short flares of 2-3 hour duration in Sector~6.

{\bf CVSO~109:} 
The two light curves look similar, with strongest power near 6 days
(6.0~days in Sector~6; 6.6 days in Sector~32). After subtracting an 8 day
running mean, the folded light curves look sinusoidal, with amplitudes of
about 0.07~mag. The autocorrelation functions are broad, consistent with a
sinusoidal modulation. 

{\bf CVSO~146:}
This star is partially blended with the brighter A2 star HD~290671 in the
$TESS$ images (4.8 pixel separation). The A star has a low amplitude
1.56 day period.
There is power at about 9 days in each sector; there is additional power at
5.5~days in Sector~6 and 4.4~days in Sector~32. None of these appear periodic.
The autocorrelation power in Sector~6 is strongest for the dips, at a period
of 10-12 days; the brightenings show a correlation at 7-8 days; in Sector~32
both the brightenings and fadings show power near 5 days, with comparable peaks
near 7 days (brightenings) and 8 days (fadings).

{\bf CVSO~165:}
$TESS$ cannot resolve this pair. The periodograms at both epochs show strongest
power at 4.3 days; in the latter half of Sector~32 there are two prominent
(0.1~mag) dips, and possibly 2 others cut off, at this spacing, but that period
is not obvious at other times. There is substructure in Sector~6, with the 
brightenings correlating with lags of 7.7 and 13 days and the fadings
correlating at 8.8~days (twice the strongest period) and 11.6 days.
In Sector~32 the brightenings and fadings correlate on the 4.3 day period.

{\bf CVSO~176:} 
This was not observed by TESS in Sector~32. In Sector~6 the character is clearly
that of a dipper, with fadings up to 0.2~mag. Despite the periodogram placing
most of the power at 3.6 days, the autocorrelation shown the strongest peak
at 7.3 days, in both brightenings and dips.

\section{Accretion column temperatures and mass flux rates} \label{sec:shock_temps}
Blackbodies are individually fit to the three accretion column spectra of each target, then a weighted average is taken by scaling the associated accretion luminosities by each column's fractional filling factor according to Equation~\ref{eq:Tshock}. Table~\ref{tab:Tshock} shows the individual temperatures fit to each accretion column.

\begin{deluxetable}{lcccc}
\tablecaption{Accretion column temperatures  \label{tab:Tshock}}
\tablehead{
\colhead{Object} & \colhead{T$_{1E10}$} & \colhead{T$_{1E11}$} & \colhead{T$_{1E12}$} & \colhead{T$_{\rm shock}$} \\
\colhead{} & \colhead{(K)} & \colhead{(K)} & \colhead{(K)} & \colhead{(K)}
}
\startdata
CVSO~58 & 4422$\pm$1 & 6880$\pm$3 & 10916$\pm$3 & 5493$\pm$96 \\
CVSO~90 & 4171$\pm$2 & 6816$\pm$3 & 10870$\pm$3 & 7975$\pm$1105 \\
CVSO~104 & 4051$\pm$2 & 6634$\pm$3 & 10698$\pm$3 & 5239$\pm$105 \\
CVSO~107 & 4187$\pm$2 & 6696$\pm$3 & 10679$\pm$3 & 5656$\pm$128 \\
CVSO~109A & 4255$\pm$2 & 6506$\pm$2 & 10714$\pm$2 & 8877$\pm$1910 \\
CVSO~146 & 4459$\pm$1 & 6871$\pm$3 & 10973$\pm$3 & 4975$\pm$27 \\
CVSO~165A & 4675$\pm$1 & 6853$\pm$3 & 10760$\pm$3 & 4800$\pm$14 \\
CVSO~165B & 4309$\pm$1 & 6690$\pm$3 & 10729$\pm$3 & 10542$\pm$343 \\
CVSO~176 & 3938$\pm$2 & 6310$\pm$1 & 10736$\pm$2 & 5043$\pm$254 \\
\enddata
\tablecomments{Best-fit temperatures for the three accretion columns and the resultant weighted-average temperature. Uncertainties on T$_{1E10}$, T$_{1E11}$, and T$_{1E12}$ are one standard deviation on the temperature. The error on T$_{\rm shock}$ comes from the propagated accretion column temperature and filling factor uncertainties. Since the filling factor uncertainties are not symmetric, the larger of the two uncertainties is used in the error propagation.}
\end{deluxetable} 

Table~\ref{tab:mass_fluxes} shows the mass flux rates of the individual accretion columns for each target. We find no clear correlations between the column that contributes the most to the total mass flux and either the stellar mass or the total mass accretion rate. Our future analysis of the entire ULLYSES sample will allow us to expand this analysis and look for correlations with stellar age.

\begin{deluxetable}{lccccc}
\tablecaption{Accretion column mass flux rates  \label{tab:mass_fluxes}}
\tablehead{
\colhead{Object} & \colhead{M$_\star$/R$_\star$} & \colhead{$\dot{M}_{1E10}$} & \colhead{$\dot{M}_{1E11}$} & \colhead{$\dot{M}_{1E12}$} & \colhead{$\dot{M}_{\rm tot}$} \\
\colhead{} & \colhead{(M$_{\odot}$/R$_{\odot}$)} & \colhead{(10$^{-8}$ \msunyr)} & \colhead{(10$^{-8}$ \msunyr)} & \colhead{(10$^{-8}$ \msunyr)} & \colhead{(10$^{-8}$ \msunyr)}
}
\startdata
CVSO~176 & 0.08 & $0.93^{+0.17}_{-0.4}$ & $0.92^{+0.22}_{-0.18}$ & $2.31^{+0.15}_{-0.13}$ & $4.18^{+0.11}_{-0.12}$ \\
CVSO~104 & 0.23 & $0.227^{+0.021}_{-0.03}$ & $0.55^{+0.04}_{-0.03}$ & $0.360^{+0.05}_{-0.022}$ & $1.14^{+0.06}_{-0.03}$ \\
CVSO~109A & 0.20 & $0.018^{+0.28}_{-0.015}$ & $2.1^{+0.7}_{-0.6}$ & $15.1^{+1.5}_{-5}$ & $17.2^{+1.0}_{-4}$ \\
CVSO~107 & 0.27 & $0.69^{+0.04}_{-0.06}$ & $0.012^{+0.23}_{-0.007}$ & $4.11^{+0.23}_{-0.4}$ & $4.83^{+0.19}_{-0.3}$ \\
CVSO~165A & 0.34 & $0.531^{+0.012}_{-0.017}$ & $0.150^{+0.020}_{-0.014}$ & $0.0253^{+0.004}_{-0.0013}$ & $0.708^{+0.020}_{-0.017}$ \\
CVSO~90 & 0.74 & $0.032^{+0.010}_{-0.019}$ & $0.0018^{+0.16}_{-0.0014}$ & $1.21^{+0.06}_{-0.6}$ & $1.25^{+0.05}_{-0.4}$ \\
CVSO~58 & 0.77 & $0.206^{+0.009}_{-0.019}$ & $0.0018^{+0.016}_{-0.0012}$ & $0.82^{+0.06}_{-0.07}$ & $1.03^{+0.06}_{-0.06}$ \\
CVSO~165B & 0.42 & $0.0007^{+0.0026}_{-0.0004}$ & $0.0053^{+0.010}_{-0.0020}$ & $1.64^{+0.05}_{-0.07}$ & $1.65^{+0.05}_{-0.07}$ \\
CVSO~146 & 0.69 & $0.220^{+0.007}_{-0.008}$ & $0.226^{+0.015}_{-0.014}$ & $0.082^{+0.007}_{-0.006}$ & $0.529^{+0.018}_{-0.016}$ \\
\enddata
\tablecomments{Mass flux rates of individual accretion columns with targets listed in order of increasing stellar mass. There is no clear correlation between the column that dominates the mass flux and either the stellar mass or the total accretion rate.}
\end{deluxetable} 

\end{document}